\documentclass[%
 reprint,
 prd,
superscriptaddress,
 amsmath,amssymb,
 aps,nofootinbib,
]{revtex4-1}
\usepackage{ulem}
\usepackage{amssymb}
\usepackage{soul}
\usepackage{dsfont}
\usepackage{graphicx}
\usepackage{dcolumn}
\usepackage{bm}
\usepackage{color}
\usepackage{epsfig}
\usepackage{hyperref}
\usepackage{natbib}
\usepackage{float}
\usepackage{footmisc}
\usepackage{lipsum}
\usepackage{amsmath}
\usepackage{braket, xcolor}
\usepackage{hyperref}
\usepackage{comment}
\usepackage{MnSymbol}
\usepackage{enumitem}
\usepackage{longtable}
\usepackage{rotating,tabularx}
\usepackage{footnote}

\hypersetup{colorlinks, citecolor=blue, linkcolor=blue, urlcolor=blue}
\usepackage[dvipsnames]{xcolor}
\usepackage[title]{appendix}

\newcommand{\Eq}[1]{Eq.~(#1)}

\newcommand{\eV}{\mathinner{\mathrm{eV}}}

\newcommand{\GeV}{\mathinner{\mathrm{GeV}}}
\newcommand{\TeV}{\mathinner{\mathrm{TeV}}}

\begin{document}

\title{Searching for heavy charged relics in the Earth}

\author{Reza~Ebadi}
\email{reza.phys@gmail.com}
\affiliation{Department of Physics and Astronomy, The Johns Hopkins University, Baltimore, MD 21218, USA}
\affiliation{Department of Physics and Astronomy, University of Delaware, Newark, DE 19716, USA}
\author{Peter~W.~Graham}
\email{pwgraham@stanford.edu}
\affiliation{Leinweber Institute for Theoretical Physics, Department of Physics, Stanford University, Stanford, CA 94305, USA}
\affiliation{Kavli Institute for Particle Astrophysics and Cosmology, Department of Physics, Stanford University, Stanford, CA 94305, USA}
\author{Surjeet~Rajendran}
\email{srajend4@jhu.edu}
\affiliation{Department of Physics and Astronomy, The Johns Hopkins University, Baltimore, MD 21218, USA}
\author{Erwin H.~Tanin}
\email{ehtanin@stanford.edu}
\affiliation{Leinweber Institute for Theoretical Physics, Department of Physics, Stanford University, Stanford, CA 94305, USA}
\author{Samuel~S.~Y.~Wong}
\email{samswong@uw.edu }
\affiliation{Leinweber Institute for Theoretical Physics, Department of Physics, Stanford University, Stanford, CA 94305, USA}
\affiliation{Department of Physics, University of Washington, Seattle, WA 98195, USA}

\begin{abstract}
We propose a method for detecting an ambient density of heavy, electrically-charged particles. Such particles would impact the Earth, lose energy in terrestrial matter, and become trapped. We study the accumulation of these rare particles in multiple target materials that provide large exposure, such as water and geological rocks. We discuss strategies for concentrating the particles by centrifugation or gravitational settling, along with particle identification using mass spectrometry. This method enables the discovery of charged relics with masses $1-10^{12}\,{\rm TeV}$ comprising a tiny fraction of the local dark matter density, reaching down to $f_X\sim 10^{-20}$ at the lowest masses. A pathfinder experiment using only a liter of water and one centrifuge (or $\sim \text{m}^3$ and no centrifuge) operating for a month can already reach $f_X\sim 10^{-10}$ and probe new parameter space.
\end{abstract}

\maketitle

\tableofcontents

\clearpage

\section{Introduction}

CHArged Massive Particles (CHAMPs) \cite{Cahn:1980ss,DeRujula:1989fe,Dimopoulos:1989hk} are generically expected in a wide range of theories beyond the standard model; see Refs.~\cite{Fairbairn:2006gg,Burdin:2014xma,Perl:2001xi} for reviews. In supersymmetric models, CHAMPs can arise either as fundamental sleptons \cite{Feng:2004mt} or as composite R-hadrons, formed when a heavy color-charged sparticle (e.g., a gluino) binds with standard model quarks to produce an electrically charged heavy particle \cite{Raby:1997pb}. CHAMPs may also appear in models with additional, vector-like fourth-generation fermions \cite{Aguilar-Saavedra:2013qpa,Greco:2014aza,Falkowski:2013jya,CMS:2024bni,CMS:2025urb}. Extra-dimensional theories predict CHAMPs in the form of Kaluza–Klein excitations \cite{Byrne:2003sa,Appelquist:2000nn}. Hidden sectors can generically host CHAMPs, again either as fundamental particles or as composite states arising from confining dark sectors \cite{Wise:2014jva,Gresham:2017zqi,Grabowska:2018lnd,Bai:2018dxf,Kaplan:2024dsn,Fedderke:2024hfy}. Even the confining sector of the standard model (QCD) can potentially host stable CHAMPs in the form of strange quark matter, also known as quark nuggets or strangelets~\cite{Bodmer:1971we,Witten:1984rs}.\footnote{These objects can be understood as superheavy nuclei containing roughly equal numbers of up, down, and strange quarks. If such quark matter exists, neutron stars are the most relevant astrophysical reservoirs, and they could be released during neutron star merger events and make their way to the Earth \cite{Witten:1984rs,DeRujula:1984axn}.}

CHAMPs can be searched for directly in particle colliders such as the Large Hadron Collider (LHC). Existing collider searches have placed strong limits on certain CHAMP realizations \cite{Gonski:2022gnz,ATLAS:2022pib,ATLAS:2022qex}. Some of us explored prospects for extending collider searches to higher masses through optimized and unconventional strategies for detecting these particles at the LHC \cite{melting_LHC_2025}. However, the mass reach for collider searches is ultimately limited by the available center-of-mass energy (a few TeV at the LHC~\cite{CMS:2024nhn}) and by rapidly suppressed production rates at high mass.

CHAMPs present in the spectrum of a theory can be produced in the late universe through high-energy phenomena such as neutron star mergers, supernovae, and cosmic-ray collisions;\footnote{These astrophysical production channels may contribute to a galactic population of CHAMPs, which is the focus of this work. Additionally, high-energy cosmic rays can directly produce CHAMPs in terrestrial matter and be detectable using our proposed methods. It would be worthwhile to calculate the production rate for this specific channel in a future work.} and in the early Universe, via thermal freeze-out \cite{Dimopoulos:1989hk} or other processes\footnote{Two illustrative examples are WIMPzillas \cite{Kolb:1998ki,Kolb:2007vd} and composite states in the QCD-like dark sectors \cite{Bai:2018dxf,Witten:1984rs}. The production mechanisms considered in these scenarios could also produce CHAMPs. For instance, gravitational production of WIMPzillas is agnostic to the particle’s electric charge and preheating after inflation can produce CHAMPs if the inflaton couples to CHAMPs \cite{Chung:1998ua,Chung:1998rq,Chung:1998bt,Kuzmin:1998kk,Hui:1998dc,Allahverdi:2002nb}. Moreover, phase transitions in dark confining sectors can generate CHAMPs in the form of dark-sector analogs of quark nuggets \cite{Bai:2018dxf,Witten:1984rs}.}. Cosmological production can access energies far above those achievable in colliders and circumvent collider mass-reach limitations. If these CHAMPs are long-lived, they can account for a fraction of the total dark matter content, and be present in the Milky Way today.

A cosmological abundance of stable CHAMPs are constrained by several complementary probes. Measurements of light-element abundances constrain CHAMPs that are present during Big Bang Nucleosynthesis (BBN)~\cite{Pospelov:2006sc,Kohri:2006cn,Kaplinghat:2006qr,Kawasaki:2007xb,Kawasaki:2007xb,Jedamzik:2007qk,Kusakabe:2017brd}.\footnote{The BBN constraint is applicable primarily to negatively-charged CHAMPS that bind to light nuclei, forming bound states with modified nuclear reaction rates and channels. Moreover, if CHAMPs are produced dominantly after BBN (e.g., through a later dark-sector phase transition), these bounds do not apply.} Some of the strongest astrophysical constraints arise from CHAMP capture and accumulation in compact objects, including neutron stars \cite{Gould:1989gw} and white dwarfs \cite{Fedderke:2019jur}, where their interactions can affect the stars' evolution and/or long-term survival. See Refs.~\cite{Cirelli:2024ssz,Burdin:2014xma,Perl:2001xi} for reviews of existing constraints on CHAMPs.

Another class of probes involves searching for anomalously heavy atoms in terrestrial matter \cite{Fleischer:1969qyz,Eberhard:1971re,Holt:1976af,Smith:1979rz,Middleton:1979zz,Smith:1982qu,Dick:1984mk,Turkevich:1984zz,Dick:1985wk,Farhi:1985ib,Nitz:1986gb,Norman:1986ux,Kovalik:1986zz,Pichard:1987ub,Norman:1988fd,Hemmick:1989ns,Blackman:1989mf,Polikanov:1990sf,Verkerk:1991jf,Boyd:1991js,Yamagata:1993jq,Vandegriff:1995ng,PerilloIsaac:1998xm,Javorsek:2001yu,Mueller:2003ji} or in extra-terrestrial samples that have landed on Earth, such as meteorites \cite{Jones:1989cq,PerilloIsaac:1998xm} and lunar rocks \cite{Stevens:1976sn,PerilloIsaac:1998xm,Han:2009sj}. Since CHAMPs follow the same chemistry as their lighter, same-charge SM counterparts \cite{Cahn:1980ss}, they may have been incorporated into terrestrial matter during Earth's formation and persist today. Alternatively, they may be captured from an ambient galactic population, losing energy in matter and accumulating over time. The sensitivity of these searches is significant only if samples of large mass and/or old age can be analyzed with high efficiency.

The strongest reported constraints arise from null results obtained through enrichment and analysis of large volumes (about $10^8$ liters) of natural water \cite{Smith:1982qu}. These samples were typically processed through commercial facilities and the original water source was not precisely identified. A key limitation that was recognized shortly after that work is that heavy CHAMPs are expected to gravitationally settle and concentrate near seabeds. As a result, a water sample from a generic depth may contain negligible amount of CHAMPs. Deep seawater samples (3–4 km below sea level) were used in later efforts to help circumvent this issue \cite{Hemmick:1989ns,Yamagata:1993jq}. However, even these searches can be subject to imperfect sampling, as CHAMPs may get trapped in deeper regions of the water pathway or in seabed rocks.\footnote{\label{ftn:other_exps}Searches using lunar rocks are, in principle, more robust. However, previous experiments either employed CHAMP identification methods based on mass spectrometry that were limited to masses $\lesssim \mathrm{TeV}$ \cite{Stevens:1976sn,Han:2009sj}, or focused on signals specific to strangelets rather than generic heavy CHAMP candidates \cite{PerilloIsaac:1998xm}. The constraints reported from searches using terrestrial samples with unknown geological histories \cite{Norman:1986ux,Norman:1988fd,Hemmick:1989ns} are difficult to interpret in terms of bounds on the galactic abundance of CHAMPs. Ref.~\cite{Javorsek:2001yu} attempted to mitigate this issue by collecting surface samples from geologically quiet sites, thereby reducing the likelihood of signal removal due to erosion or other geological processes. Nevertheless, even in such regions, the surface-level geological systematics remain substantial and limit the robustness of the inferred constraints. Ground- and underground-based experiments \cite{Kajino:1984ug,MACRO:2002jdv,Majorana:2018gib} are insensitive to lighter ($\lesssim 10^4\TeV$) CHAMPs because these CHAMPs stop in the atmosphere, neutralize by binding with an electron or a nucleus, and thermalize to low velocities before reaching the detector \cite{Burdin:2014xma,Perl:2001xi}.}

\begin{figure*}[t!]
    \centering
    \includegraphics[width=0.325\linewidth]{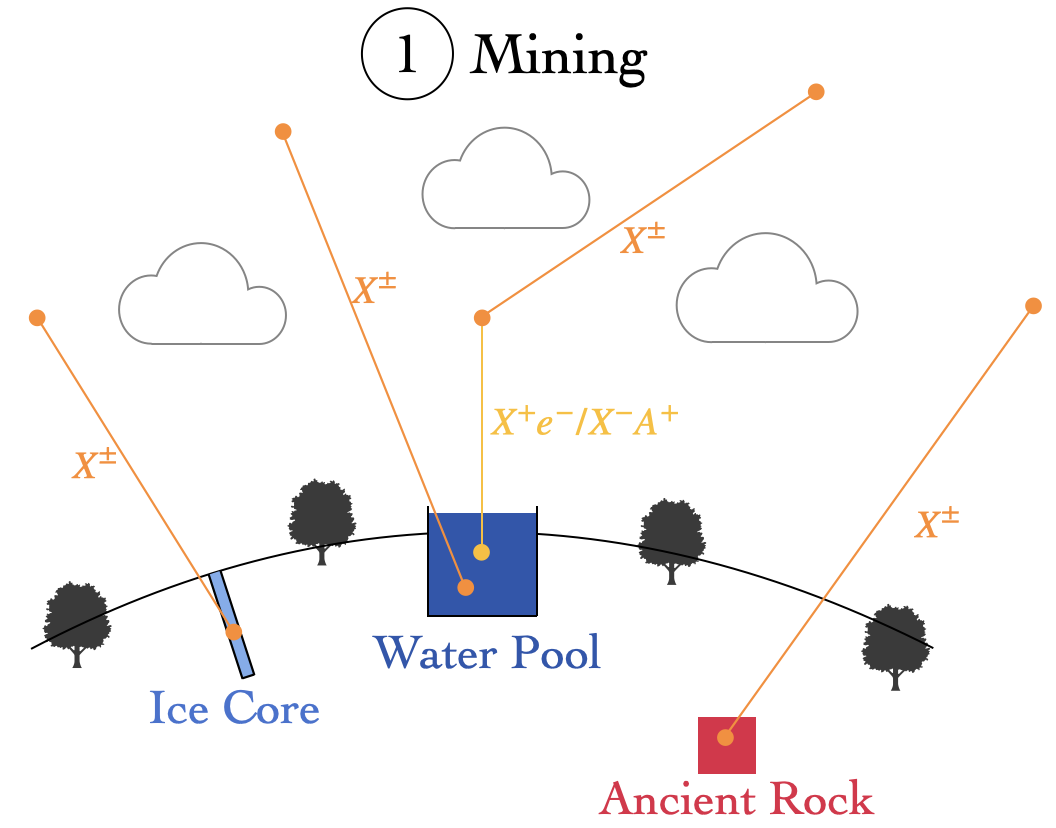}
    \includegraphics[width=0.665\linewidth]{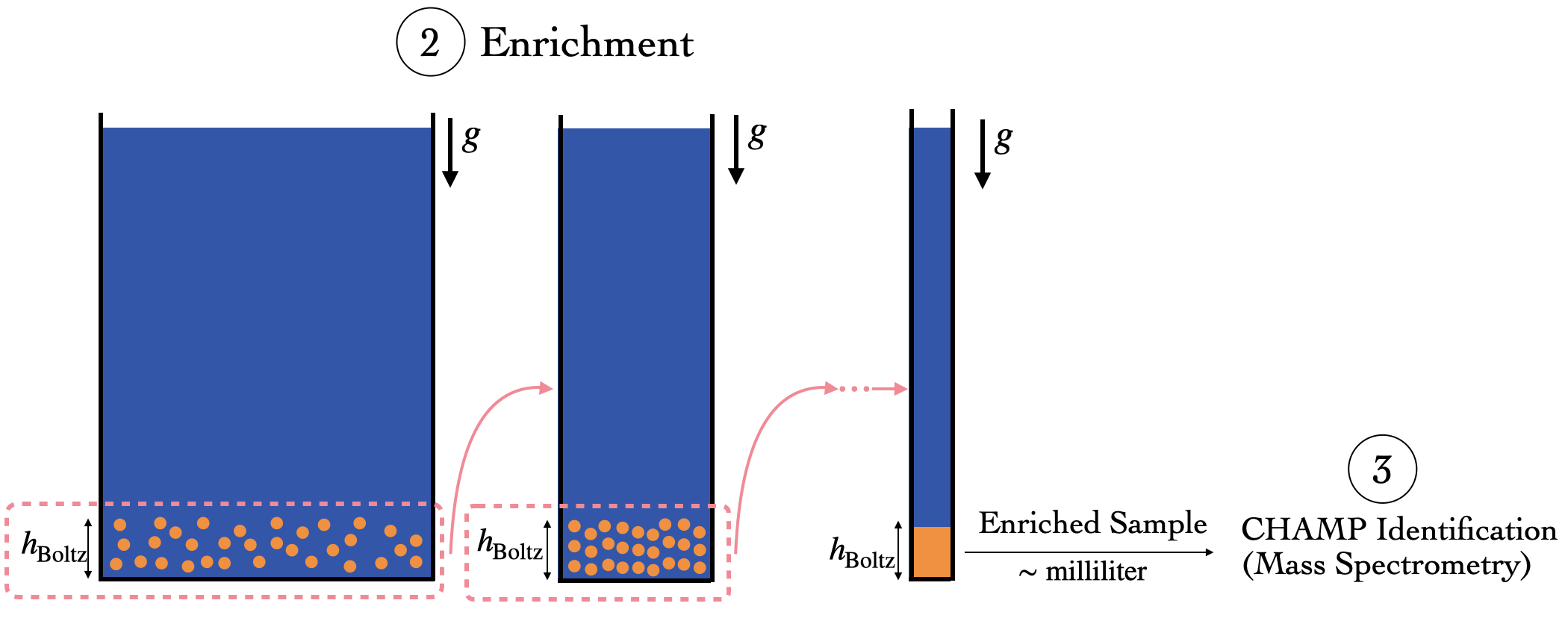}
    \caption{Schematic illustration of the proposed experimental strategy.}
    \label{fig:scheme}
\end{figure*}

In this work, we present robust and scalable strategies to search for galactic CHAMP relics and estimate the resulting experimental reach. The proposed experimental protocol is summarized in Fig.~\ref{fig:scheme} and consists of three steps: CHAMP accumulation in artificial or natural stopping targets followed by extraction of relevant materials (mining), increasing the CHAMP concentration by reducing the sample volume while minimizing CHAMP losses (enrichment), and direct identification of CHAMPs using, e.g.,  mass-spectrometric techniques that can  discover or exclude particles with anomalously low charge-to-mass ratios (detection). The main objectives and results of this work are:

\textit{CHAMP accumulation.---} Interpreting terrestrial search results requires a well-founded estimate of where incoming CHAMPs would stop and accumulate in the Earth. To the best of our knowledge, such a study has not previously been carried out. We take a first step by computing the distribution of stopping points for galactic, virialized CHAMPs impinging on the Earth as they lose energy and come to rest in matter. The details are presented in Section~\ref{sec:accumulation}, with additional technical derivations provided in Appendixes~\ref{app:stoppingdist} and \ref{app:eqtime}.

\textit{Target samples.---} The obtained stopping distribution informs the optimal choice of stopping targets. Low-mass CHAMPs tend to stop near the surface, motivating shallow targets whose exposure history is known and monitored. Heavier CHAMPs have lower incoming fluxes and stop deeper, motivating ancient (i.e., long exposure time) natural samples drawn from greater depths.
These considerations and the target selection are discussed in Section~\ref{sec:target}.

\textit{Sample processing and CHAMP detection.---} Any competitive search must process large amount of materials, requiring enrichment to concentrate any accumulated CHAMPs into a tractable volume for high-precision CHAMP identification methods. We show that efficient enrichment is possible over much of the parameter space by exploiting gravitational settling of heavy CHAMPs in liquid samples  (``gravitational enrichment"), with centrifugal enrichment \cite{melting_LHC_2025} providing a complementary approach in regimes where sinking is ineffective. The details are presented in Section~\ref{sec:detection}, where we also discuss prospects for precision charge-to-mass ratio measurements that can discriminate CHAMPs from backgrounds.

The projected sensitivity of these strategies is shown in Fig.~\ref{fig:reach},  which summarizes the reach in the parameter space of galactic CHAMP abundance versus CHAMP mass.

\section{CHAMP accumulation in the Earth}
\label{sec:accumulation}

\subsection{Assumptions}

CHAMPs, being electrically charged, are expected to undergo complex phase-space dynamics within the Milky Way. In particular, their trajectories and spatial distribution are influenced by astrophysical electromagnetic fields. The distribution of CHAMPs in the Milky Way has been the subject of scrutiny over the past decades. It was previously argued that CHAMPs may be expelled from the galactic disk by supernova activity \cite{Dimopoulos:1989hk} and prevented from reentering by galactic magnetic fields \cite{Chuzhoy:2008zy}. However, recent studies with more realistic magnetic-field morphologies challenge this picture~\cite{Perri:2025tvq}.\footnote{There may also exist an accelerated population of CHAMPs with a velocity distribution different from the standard virial distribution \cite{Dunsky:2018mqs}.}

We do not attempt to model the detailed cosmology and astrophysics of CHAMPs here. Instead, we focus on probing a local CHAMP density near the Earth, independently of its origin. We assume that CHAMP particles $X^\pm$ of charge $\pm e$ and mass $m_X$ exist in the galaxy with a local mass density $\rho_X=f_X\rho_{\rm DM}$, where $f_X\ll 1$ and $\rho_{\rm DM}=0.4\GeV/\text{cm}^3$ is the dark matter density near Earth. Such $X$ particles impact the Earth with a total flux of
\begin{align}
    \Phi_X= f_X\frac{\rho_{\rm DM}}{m_X}v_{\rm DM} ~,
    \label{eq:flux}
\end{align}
where $v_{\rm DM}=10^{-3}$ is roughly the typical virial velocity in our galaxy. 

We further assume the $X$ particles arrive on Earth with velocities $v_i$ following an isotropic Maxwell-Boltzmann distribution
\begin{align}\label{eq:speeddist}
    f(v_i)=\frac{4\pi v_i^2e^{-v_i^2/2\sigma_v^2}}{(2\pi)^{3/2}\sigma_v^3} ~,
\end{align}
with a velocity dispersion $\sigma_v=10^{-3}$. Note that CHAMPs with mass $m_X \gtrsim 1\,\TeV$ moving with velocities $v \sim 10^{-3}$ have rigidities $p/e \gtrsim 1\,\GeV$, and are therefore not subject to significant deflection by the solar wind magnetic field \cite{Cholis:2015gna,Potgieter:2013mcc} or the Earth’s magnetic field. We will henceforth neglect magnetic forces on CHAMPs.

We are mainly interested in detecting CHAMPs in the mass range $m_X=1-10^{14}\TeV$. CHAMPs lighter than $\sim \TeV$ can be produced in colliders and are essentially ruled out. CHAMPs heavier than $\sim 10^{14}\TeV$ would sink rapidly to the center of the Earth due to their weight alone and are not stopped by solid material. This is because the gravitational potential energy drop of such a CHAMP when it sinks by $\sim 1 \text{\AA}$ exceeds $\sim 1\eV$, the typical energy barrier for breaking chemical bonds or necessary rearrangements of surrounding atoms and molecules.

CHAMPs with incoming speeds $v\lesssim v_{\rm esc,\oplus}=4\times 10^{-5}$ are significantly affected by Earth's gravity, making the distribution of stopping points for such slow CHAMPs difficult to estimate. We conservatively neglect this low-velocity subpopulation. For $m_X\gtrsim 3\times10^{12}\TeV$ only particles in this low-velocity tail can stop in the Earth (see Appendix~\ref{app:stoppingdist}). Accordingly, we limit our analysis to $m_X\lesssim 3\times10^{12}\TeV$. We also note that our assumed CHAMP speed distribution, Eq.~\eqref{eq:speeddist}, neglects Earth's velocity boost and the galactic escape-speed cutoff at Earth's location ($\sim 540\text{ km/s}$). We will comment later how these assumptions affect the projected reach.

\subsection{Distribution of stopping points}
Depending on the mass of the CHAMP particle, it will \textit{stop} at different locations within the Earth or the Earth's atmosphere. Upon stopping, its kinetic energy becomes sufficiently low to capture an oppositely charged particle and form a neutral bound state. In this subsection, we study the stopping process of CHAMPs, while in the next subsection we study the subsequent dynamics of the CHAMPs towards equilibrium in their neutral bound state, and in the presence of gravity.

We model the Earth as a uniform sphere of density\footnote{The central densities of the Earth are much higher. However, in practice we can access only the first $\sim{\rm km}$ of the Earth's surface, so only CHAMPs stopping at such shallow depths are relevant, thus justifying this approximation.} $\rho_{\oplus} = 2.5\,\mathrm{g/cm^3}$ and radius $R_\oplus = 6378\,\mathrm{km}$, covered by an atmosphere with mass column density $\sim P_{\rm sea}/g= 1\,\mathrm{kg/cm^2}$, where $P_{\rm sea} = 10^{5}\,\mathrm{Pa}$ is the atmospheric pressure at sea level and $g_ = 10\,\mathrm{m/s^2}$ is the surface gravitational acceleration. The atmospheric column density is comparable to that of typical Earth rocks of thickness
\begin{align}\label{eq:solid_equiv_hatm}
L_{\rm atm}=\frac{P_{\rm sea}/g}{\rho_\oplus} \simeq 4\,{\rm m} ~.
\end{align}
Below the atmosphere lies the {\it lithosphere}, a solid crust of thickness $L_{\rm litho}\approx 100\,\mathrm{km}$, beneath which rocks begin to melt. 

The lithosphere acts as a stopping target for incoming $X$ particles. Those that are stopped in the lithosphere accumulate over time, whereas those that stop below it are effectively lost. Within the lithosphere, heavier particles that are stopped in the rocks are, for the most part, expected to remain bound to the surrounding material. Lighter $X$ particles that come to rest in a fluid environment, i.e., the atmosphere or the ocean, eventually thermalize with the surrounding medium and are subsequently distributed according to the Boltzmann distribution, as discussed in the next subsection and Appendix~\ref{app:eqtime}.

\begin{figure}[t!]
    \centering
    \includegraphics[width=\linewidth]{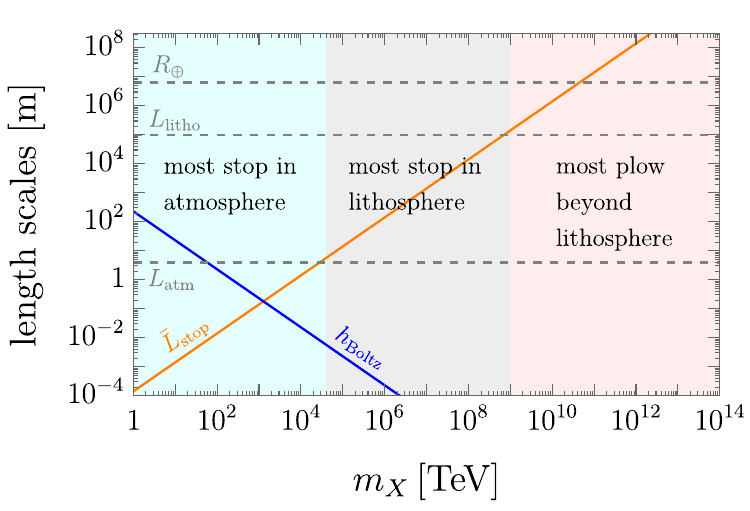}
    \caption{The typical stopping length $\overline{L}_{\rm stop}$ [orange solid, Eq.~\eqref{eq:lbarstop}] and Boltzmann height $h_{\rm Boltz}$ [blue solid, Eq.~\eqref{eq:hBoltz}] of CHAMPs as functions of their mass $m_X$. The Earth’s radius $R_\oplus$, lithosphere thickness $L_{\rm litho}=100\text{ km}$, and the characteristic rock-equivalent atmospheric height $L_{\rm atm}$ [Eq.~\eqref{eq:solid_equiv_hatm}] are shown in gray dashed lines for comparison. }
    \label{fig:lenghtscales}
\end{figure}

As shown in Appendix~\ref{app:stoppingdist}, the stopping length in terrestrial matter for an $X$ particle with initial velocity $v_i \lesssim 10^{-3}$ is
\begin{align}
    L_{\rm stop}&\approx 0.1~\text{mm}\left(\frac{v_i}{10^{-3}}\right)\left(\frac{m_X}{1~\TeV}\right)\left(\frac{\rho_{\rm matter}}{2.5~\text{g}/\text{cm}^3}\right)^{-1} ~.
    \label{eq:Lstop}
\end{align}
We define the typical {\it rock-equivalent stopping length} as
\begin{align}
    \bar{L}_{\rm stop}&\equiv L_{\rm stop}\left(v_i=\sqrt{2}\sigma_v,\rho_{\rm matter}=\rho_\oplus\right)~.
    \label{eq:lbarstop}
\end{align} 
We can roughly delineate three mass regimes based on where the typical $X$ particles come to a \textit{first} stop (see Fig.~\ref{fig:lenghtscales}):
\begin{itemize}
    \item \textit{Most stop in atmosphere} ($m_X\lesssim 10^{4}\TeV$): In this regime, $\bar{L}_{\rm stop} \lesssim L_{\rm atm}$, so a typical $X$ particle slows down to thermal speeds within the atmosphere.
    \item \textit{Most stop in lithosphere} ($m_X\sim 10^{4}\text{--}10^9\TeV$): This regime corresponds to $\bar{L}_{\rm stop} \sim 1\text{--}10^{5}\,\mathrm{m}$, corresponding to $X$ particles coming to rest below the atmosphere and within the lithosphere.
    \item \textit{Most plow beyond lithosphere} ($m_X\sim 10^{9}\text{--}10^{14}\TeV$): In this mass range, $\bar{L}_{\rm stop} \sim 10^{5}\text{--}10^{11}\,\mathrm{m}$, so a typical $X$ particle reaches the liquid layers of the Earth or passes through it entirely. In this regime, any accumulated CHAMPs come either from the low-velocity tail of the incoming CHAMP velocity distribution [Eq.~\eqref{eq:speeddist}], which is only power-law suppressed, or from particles that traverse the Earth and lose enough energy to stop in the lithosphere on the far side. As noted above, for $m_X\gtrsim3\times 10^{12}\TeV$ the CHAMPs that can be stopped are those arriving with $v\lesssim v_{\rm esc,\oplus}$ for which Earth's gravity is significant. Because our analysis neglects Earth's gravity, its validity is limited to $m_X\lesssim 3\times 10^{12}\TeV$.     
\end{itemize}

\begin{figure}[t!]
    \centering
    \includegraphics[width=\linewidth]{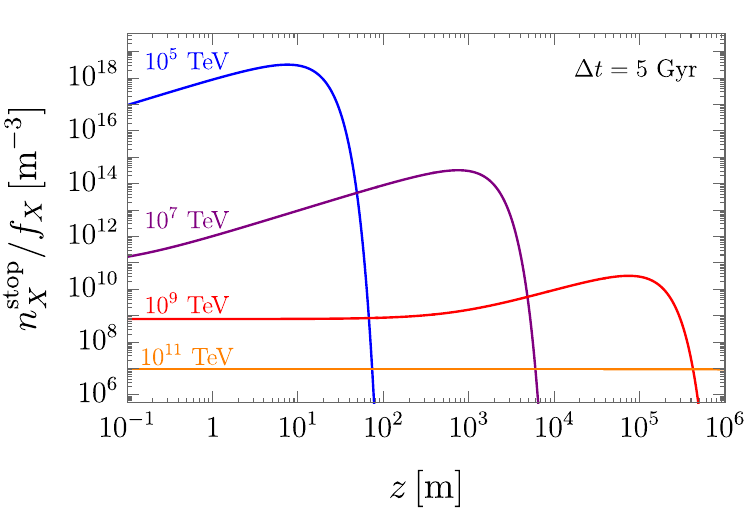}
    \caption{Density of \textit{first} stopping points of $X$ as a function of rock-equivalent depth $z$ for $m_X=10^{5}, 10^{7}, 10^{9}, 10^{11}\TeV$. We set $\sigma_v = 10^{-3}$ and $\Delta t = 5\,{\rm Gyr}$. Since $n_X^{\rm stop} \propto  \Delta t$, the results for other values of $\Delta t$ can be obtained by trivial rescaling. 
    }
    \label{fig:nxvsz}
\end{figure}

Given the entry point, incoming direction, and stopping length $L_{\rm stop}(v_i)$ of an $X$ particle, its stopping location can be determined through straightforward geometric considerations. Over a time interval $\Delta t$, the number density of $X$ particles accumulated at a depth $z$ satisfying $0 \leq z \ll R_\oplus$ is given by (see Appendix~\ref{app:stoppingdist})
\begin{align}
    n_X^{\rm stop}(z)=&\frac{\Phi_X\Delta t}{4R_\oplus}\left\{\left[\text{Erf}\left(\frac{2R_\oplus}{\bar{L}_{\rm stop}}\right)-\text{Erf}\left(\frac{z}{\bar{L}_{\rm stop}}\right)\right]\left(1+\frac{4R_\oplus z}{\bar{L}_{\rm stop}^2}\right)\right.\nonumber\\
    &\left.+\frac{2}{\sqrt{\pi}}\left(\frac{2R_\oplus}{\bar{L}_{\rm stop}}e^{-\frac{4R_\oplus^2}{\bar{L}_{\rm stop}^2}}-\frac{z}{\bar{L}_{\rm stop}}e^{-\frac{z^2}{\bar{L}_{\rm stop}^2}}\right)\right\}~,
    \label{eq:nXstop}
\end{align}
for $z/\bar{L}_{\rm stop}\gg v_{\rm esc,\oplus}/(\sqrt{2}\sigma_v)=0.03$. The quantities $\Phi_X$ and $\bar{L}_{\rm stop}$ are given in Eq.~\eqref{eq:flux} and Eq.~\eqref{eq:lbarstop}, respectively. In Fig.~\ref{fig:nxvsz}, we show $n_X^{\rm stop}/f_X$ as a function of $z$ for $\Delta t=5~\text{Gyr}$ and $m_X=10^{5}, 10^{7}, 10^{9}, 10^{11}\TeV$. The general trend is that as the mass $m_X$ increases, the incoming flux decreases as $\Phi_X \propto m_X^{-1}$, while the typical $X$ particles penetrate deeper into the Earth; this results in a smaller number density at shallow depths, but a distribution that reaches larger penetration depths.

In the limit of $z,\bar{L}_{\rm stop}\ll R_\oplus$, the $n_X^{\rm stop}$ simplifies to 
\begin{align}
    n_X^{\rm stop}\approx \frac{\Phi_X\Delta t}{4R_\oplus}\left[1-\text{Erf}\left(\frac{z}{\bar{L}_{\rm stop}}\right)\right]\left(1+\frac{4R_\oplus z}{\bar{L}_{\rm stop}^2}\right) ~.
    \label{eq:nXstopshortLstop}
\end{align}
In this regime, $n_X^{\rm stop}(z)$ is approximately flat for $z\lesssim \bar{L}_{\rm stop}^2/4R_\oplus$. Then $n_X^{\rm stop}(z)$ increases roughly linearly with $z$ for $\bar{L}_{\rm stop}^2/4R\oplus \lesssim z \lesssim \bar{L}_{\rm stop}$ (where the flat-Earth approximation holds), peaks at $z \sim \bar{L}_{\rm stop}$, and is rapidly suppressed for $z \gtrsim \bar{L}_{\rm stop}$. These behaviors are shown by the blue ($m_X=10^5\TeV$), purple ($m_X=10^7\TeV$), and red ($m_X=10^9\TeV$) curves in Fig.~\ref{fig:nxvsz}.

In the limit $z\ll R_\oplus\ll \bar{L}_{\rm stop}$, only $X$ particles from the low-velocity tail ($v_i \ll \sigma_v$) of the initial Maxwell–Boltzmann distribution are stopped in the Earth. In this case, $n_X^{\rm stop}$ is approximately uniform throughout the Earth,
\begin{align}
    n_X^{\rm stop}\approx \frac{2\Phi_X\Delta t}{\sqrt{\pi}\bar{L}_{\rm stop}}  ~.
    \label{eq:nXstoplongLstop}
\end{align}
The orange curve ($m_\chi=10^{11}\,{\rm TeV}$) in Fig.~\ref{fig:nxvsz} reflects this flat distribution.\footnote{The stopping-point distribution derived here may also be relevant for other experiments searching for damage tracks in rocks \cite{Ebadi:2021cte,Baum:2023cct,Hirose:2025jht}. Since the damage tracks left by particles are generally velocity dependent, tracks formed near their stopping points are expected to have distinctive features.}

\subsection{Redistribution after stopping}
CHAMPs that come to a stop in a solid (e.g., rock  or ice) will remain localized at their stopping points for the same reasons atoms in a solid do not move around in general. In contrast, CHAMPs that stop in a fluid (e.g., air or water), undergo further post-stopping dynamics: capture of SM charges, diffusion, gravitational drift, and interactions with solid boundaries, tending to relax toward a Boltzmann equilibrium, with a number density profile 
\begin{align}
    n_X(h)=n_{X,0} \, e^{-h/h_{\rm Boltz}} ~,
\end{align}
where $h$ is the height above a solid interface, and the Boltzmann height $h_{\rm Boltz}\equiv T/(m_X g)$ is 
\begin{align}\label{eq:hBoltz}
    h_{\rm Boltz}&=230~\text{m}\left(\frac{m_X}{1~\TeV}\right)^{-1}\left(\frac{g}{10~\text{m}/\text{s}^2}\right)^{-1} ~,
\end{align}
at room temperature $T=300~\text{K}$. We estimate the timescale to reach this equilibrium in Appendix~\ref{app:eqtime}. Here we summarize the qualitative relaxation processes and their parametric timescales.

After slowing to sufficiently low kinetic energy through Lindhard-Scharff stopping associated with its electric charge, a CHAMP $X^\pm$ captures an electron or an ion $A^+$, and forms a neutral bound state, $X^+e^-$ or $X^-A^+$. Once neutralized, the bound state no longer experiences the same Lindhard-Scharff stopping that slowed the original charged particle, and its subsequent motion in a fluid is controlled by gravity and collisions with atoms and molecules of the medium.

For concreteness, we focus on relaxation in water for neutralized CHAMP bound states starting at a typical height $H=10~\text{m}\gg h_{\rm Boltz}$, in which case Boltzmann equilibration requires downward redistribution. For $m_X\lesssim 4\times 10^6\,\TeV$, the heavy particle undergoes a random walk with a small downward drift due to gravity. The approach to equilibrium is controlled by gravitational drift (sedimentation) and occurs in a timescale of $t_{\rm drift}\sim 0.6\,\text{yr}\,(m_X/100\TeV)^{-1}$. We find that these $m_X\lesssim 4\times 10^{6}\TeV$ particles bounce off a solid interface instead of plowing through it.

For larger masses, $m_X\gtrsim 4\times 10^6\TeV$, the biased-diffusion picture breaks down and the particle instead undergoes direct fall through the fluid, reaching the bottom solid interface on timescales of $\lesssim \text{minutes}$. After reaching the bottom, the bound state may either bounce and remain in the fluid, or penetrate into the solid.

For intermediate masses $m_X\sim 4\times 10^6-2\times 10^7$~TeV, the heavy particle typically undergoes repeated bounces while drag in the fluid dissipates its kinetic energy, thermalizing it on timescales of $\mathcal{O}(\text{minutes})$, set by its initial fall.

For sufficiently large masses $m_X\gtrsim 2\times 10^7$~TeV, the heavy particle plows into the solid and stops after a microscopic distance $\lesssim 10\mu\text{m}$. This motivates using a recoverable capture layer, such as a tin lining, at the bottom of a fluid reservoir to prevent loss of $m_X\gtrsim 2\times 10^7\TeV$ CHAMPs into container-bottom walls.

\section{Mining}
\label{sec:target}

\begin{figure*}[p]
    \centering
    \includegraphics[width=\linewidth]{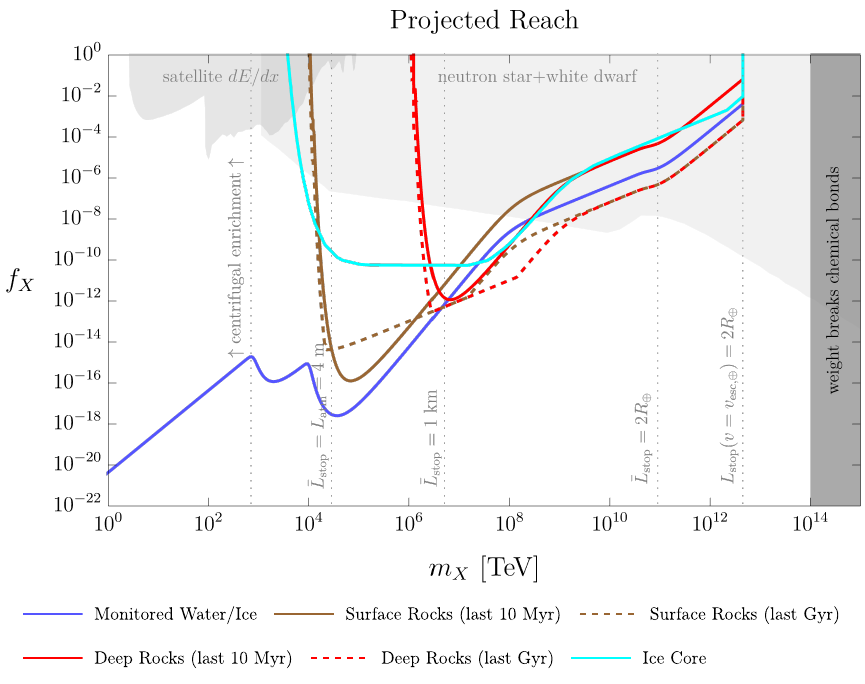}
    \caption{Projected sensitivity of our proposed searches to singly charged CHAMPs $X^\pm$ with mass $m_X$ and galactic density $\rho_X=f_X\times 0.4\GeV/\text{cm}^3$. For monitored water/ice, we assume $t_{\rm exp}=10~\text{yr}$ and $V_{\rm exp}=10^{5}~\text{m}^3$ for $m_X\lesssim 10^3\TeV$ (limited by centrifugal enrichment) and $V_{\rm exp}=4\times 10^{7}~\text{m}^3$ for $m_X\gtrsim  10^3\TeV$ (improved by gravitational enrichment). The transition between the two is smoothed over. For rocks, we assume a fiducial volume $1~\text{m}^3$ and depths 10 m (surface rocks) or 1 km  (deep rocks). We show the reaches considering  the rock's exposure over the last 10 Myr (approximately the timescale of geological stirring) during which we expect the sample to remain approximately at the depth where it is found as well as the last Gyr during which we assume the sample's depth to be approximately uniformly distributed over a geological smearing length scale of $L_{\rm geo}\sim 100~\text{km}$. For ice core, we assume sample dimensions $t_{*}\times z_{\rm max}\times A_{\rm core}=0.25~\text{Myr}\times 3~\text{km}\times 10^{-2}~\text{m}^2$. The optimal sensitivity range in our proposal is complementary to the higher-mass regime probed by neutron-star and white-dwarf stability considerations \cite{Gould:1989gw,Fedderke:2019jur}. Note that the combined neutron star+white dwarf limits for $X^+$ and $X^-$ are nearly identical; 
    here we show the very slightly stronger $X^+$ limit. The ``satellite $dE/dx$" constraint is derived from a reanalysis of data collected by the solid-state detectors and plastic scintillators aboard the Interplanetary Monitoring Platform-8 (IMP-8) mission, as reported in Ref.~\cite{snowden1990search},\protect\footnotemark[1] and is, to our knowledge, the most robust limits currently available in the literature in the $m_X\lesssim 10^4\TeV$ regime.\protect\footnotemark[2]}

\protect\footnotetext[1]{\raggedright Note that the $dE/dx$ of CHAMPs assumed in Ref.~\cite{snowden1990search} is different from what we assume here~[\Eq{\ref{eq:dEdx}}], which is based on the muon stopping range reported in Ref.~\cite{ParticleDataGroup:2024cfk}.}
\protect\footnotetext[2]{\raggedright Other bounds derived from searches in terrestrial matter typically rely on samples with poorly known pre-search histories, rendering the corresponding constraints on the galactic abundance uncertain and difficult to interpret reliably. In the case of well-characterized lunar soil samples, the detection methods are sensitive primarily to strangelets rather than to $X^{\pm}$ particles \cite{PerilloIsaac:1998xm} (see footnote~\ref{ftn:other_exps}).}
    
    \label{fig:reach}
\end{figure*}

The sensitivity of our proposed experiment is determined by several target-dependent factors: the depth at which the material has accumulated CHAMPs, the duration of exposure to the CHAMP flux, the total target area and/or volume that can be obtained and processed, and compatibility of the sample material with enrichment methods.

Fig.~\ref{fig:nxvsz} shows that the stopping depth for a given CHAMP depends on its mass $m_X$, peaking at $z \sim \bar{L}_{\rm stop} \propto m_X$. Thus, samples extracted from different depths probe different CHAMP mass ranges, with deeper samples offering enhanced sensitivity to heavier CHAMPs. In practice, obtaining samples from depths beyond about a $\sim \mathrm{km}$ within the Earth is highly challenging, which weakens the reach of our proposal for masses above $m_X \sim 2\times 10^{7}\,\mathrm{TeV}$ (Fig.~\ref{fig:nxvsz}). In this higher-mass regime, geological rock samples extracted from the deepest accessible regions of the Earth are particularly advantageous (Fig.~\ref{fig:reach}).

Motivated by these considerations, we propose two complementary categories of target materials: (i) monitored water and/or ice, and (ii) ancient samples, including geological rocks and ice cores. The projected sensitivities for representative benchmark implementations of these search strategies are shown in Fig.~\ref{fig:reach}, and we discuss each in detail below.

In estimating this (ultimate) reach, we assume a background-free search with $\mathcal{O}(1)$ efficiency; that is, all the CHAMP particles present in the entire initial large-volume sample are successfully retained during enrichment and subsequently identified in the detection stage, while any heavy standard model molecules (or other backgrounds) are distinguishable from CHAMPs. The projected sensitivity is obtained by requiring that the number of CHAMP particles captured in the \textit{initial} (pre-enrichment) sample is
\begin{align}
    N_X^{\rm sample}\gtrsim 1 ~.
\end{align}
We discuss in Section~\ref{sec:detection} and Appendix~\ref{app:enrichment} why this assumption is reasonable. 
Notice that varying this criterion by an $\mathcal{O}(1)$ factor $c$ (i.e.,~demanding $N_X^{\rm sample}\gtrsim c$) shifts the inferred reach in $f_X$ by the same factor $c$. Since our projection spans many orders of magnitude in $f_X$, this choice has a minor impact on our results.

Furthermore, as noted earlier, we assume a pure Maxwellian speed distribution of arriving CHAMPs as shown in Eq.~\eqref{eq:speeddist}. This approximation neglects several effects that can modify the detailed shape of the velocity distribution, including the galactic escape-speed cutoff, possible magnetic-field effects, as well as Earth's gravitational pull and the boost from the Earth's motion relative to the halo. We expect these refinements to move the $f_X$ sensitivity and the low-$m_X$ cutoff by at most an $\mathcal{O}(1)$ factor.

\subsection{Monitored samples: water and ice}

Near-surface solids at depths $\lesssim 10~\text{m}$ are vulnerable to various erosion effects (earthquakes, flooding, dinosaur footsteps, etc.) which could remove or randomly redistribute any accumulated CHAMPs. Since lighter CHAMPs tend to accumulate at shallow depths, a robust search must use targets whose near-surface layers remain pristine and whose exposure history is known. The clean way to achieve this is to prepare the target ourselves and monitor it over the accumulation period. This avoids the history-dependent systematics that complicate earlier searches, e.g., those based on natural water \cite{Smith:1982qu} and other poorly characterized samples.

For $m_X\lesssim 10^4\TeV$, most of the $X^\pm$ particles are expected to stop in the atmosphere. A fluid target exposed to air can then capture them efficiently after neutralization, as they sediment into the fluid rather than being lost to unprotected surface solids  or dispersed by environmental transport.\footnote{We do not expect water surface tension to provide a meaningful barrier for an \AA-size neutralized $X$ bound state. The reason is that liquid-vapor interface of water is dynamically fluctuating. Near-surface molecules continually break (evaporation) and reform (condensation) bonds. Consequently, even extremely slow $X$ bound state can enter water.}

We therefore propose two complementary types of monitored targets: a large-volume water reservoir (``water pool") and natural ice. A water pool is particularly effective for low-mass CHAMPs that stop in the atmosphere and subsequently sink; it is also already in the liquid phase, which enables immediate enrichment.

Natural ice enables large target volumes without having to construct a dedicated reservoir and it can be monitored to ensure shallow layers have not melted or been disturbed. The trade-off is that ice requires a controlled melting step prior to enrichment. Overall, monitored water and ice targets are more scalable and more straightforward to enrich compared to, e.g., geological rocks, which we discuss in the next subsection.

Let $V_{\rm pool}$ be the volume of the pool, $h_{\rm pool}$ its depth, and $t_{\rm exp}$ its exposure time. Since water is about $2.5$ times less dense than the density of a typical rock, the rock-equivalent depth of the pool is $L_{\rm pool}\sim h_{\rm pool}/2.5$. As fiducial parameters, we take $h_{\rm pool}=10~\text{m}$ and $t_{\rm exp}=10~\text{yr}$. CHAMPs that stop in the pool stay in the pool because diffusion out of the water is slow for our benchmark  parameters. In addition, CHAMPs that stop in the atmosphere subsequently drift/diffuse into the pool (see Fig.~\ref{fig:scheme}), providing an important extra capture channel at low masses.

The total number of CHAMPs accumulated in the pool can be written as a sum of CHAMPs that stop directly in the water and CHAMPs supplied from the atmosphere,
\begin{align}
    N_X^{\rm pool}=&
   V_{\rm pool} \frac{\int_{L_{\rm atm}}^{L_{\rm atm}+L_{\rm pool}} dz\,n_X^{\rm stop}(\Delta t=t_{\rm exp},z) }{h_{\rm pool}}+V_{\rm pool} n_X^{\rm air} ~,
\end{align}
where $L_{\rm atm}\approx 4~\text{m}$ [Eq.~\eqref{eq:solid_equiv_hatm}], $n_X^{\rm stop}$ is the number density of CHAMPs at first-stopping points as a function of (rock-equivalent) depth $z$ previously found in Eq.~\eqref{eq:nXstop}, and $n_X^{\rm air}$ is the number density of airborne neutralized CHAMP bound states.

To estimate $n_X^{\rm air}$, we treat the lower atmosphere as effectively homogenized over a characteristic mixing height $h_{\rm atm}\sim 10~\text{km}$ by atmospheric stirring which typically occurs on $\sim$ days-weeks timescale \cite{seinfeld2016atmospheric}, i.e., faster than the timescale for neutralized CHAMPs to drift or diffuse across $h_{\rm atm}$. We assume airborne CHAMPs are depleted primarily by loss into oceans on a timescale set by gravitational drift to traverse $\sim h_{\rm atm}$ after the particle enters ocean water, $t_{\text{atm}\rightarrow\text{sea}}\sim t_{\rm drift}(\rho_A=1~\text{g/cm}^3, H\sim h_{\rm atm})\sim 600~\text{yr}\,(m_X/100\TeV)^{-1}$, where the drift time $t_{\rm drift}$ is estimated in Appendix~\ref{app:eqtime} [\Eq{\ref{eqn:drifttime}}] and $\rho_A$ is the water density. This is the timescale for $\sim h_{\rm atm}$ worth of airborne neutralized CHAMPs per unit area to sink into the ocean and be removed from the atmosphere. 
This depletion time sets the steady-state airborne CHAMP density,
\begin{align}
    n_X^{\rm air}\sim \frac{1}{h_{\rm atm}}\int_{0}^{L_{\rm atm}}dz\,n_X^{\rm stop}(\Delta t= t_{\text{atm}\rightarrow\text{sea}},z) ~,
\end{align}
where we divided the column density of atmosphere-stopped $X$ by $h_{\rm atm}$ due to the homogenizing effect of atmospheric stirring. For $m_X\lesssim 10^{4}\TeV$, the stopping distribution in the atmosphere scales as $n_X^{\rm stop}\propto \Phi_X t_{\rm atm\rightarrow sea}$. Since $\Phi_X\propto m_X^{-1}$ and $t_{\rm atm\rightarrow sea}\propto m_X^{-1}$, this implies $n_X^{\rm air}\propto m_X^{-2}$.

In Fig.~\ref{fig:reach}, we show the projected sensitivity for monitored water/ice with exposure time $t_{\rm exp}=10~\text{yr}$, height $h_{\rm pool}=10~\text{m}$, and total volume $V_{\rm exp}=10^5~\text{m}^3$ (limited by centrifugal enrichment) for $m_X\lesssim 10^{3}\TeV$  and $V_{\rm exp}=4\times 10^7~\text{m}^3$ (improved by gravitational enrichment) for $m_X\gtrsim 10^3\TeV$; see Section~\ref{sec:detection} and Appendix~\ref{app:icecore} for more details behind the choice of $V_{\rm exp}$.

We do not differentiate between a monitored water pool and monitored natural ice in Fig.~\ref{fig:reach}. In practice, one may choose between them depending on logistics and attainable volumes. For example, at very high masses, extremely large effective volumes may be easier to realize with monitored ice, whereas water pools are particularly advantageous at low masses because they can capture atmosphere-stopped CHAMPs. Our projected sensitivity spans many orders of magnitude in both CHAMP mass and galactic abundance, going beyond existing bounds at masses $m_X\lesssim 10^{9}\,{\rm TeV}$. This regime is complementary to the stringent existing constraints derived from neutron star and white dwarf observations \cite{Gould:1989gw,Fedderke:2019jur}.

The main features of the reach curve can be understood as follows. For $m_X \lesssim 10^4\,\TeV$, essentially all incoming $X$ particles within the area of the water pool are stopped in the atmosphere and subsequently contribute to the pool, resulting in $N_X^{\rm pool} \propto n_X^{\rm air}\propto m_X^{-2}$. Around $m_X\sim 10^3\TeV$, gravitational enrichment (Section~\ref{sec:detection}) becomes feasible, boosting the $f_X$ sensitivity for $m_X\gtrsim 10^3\TeV$, as reflected by the sharp feature at $m_X\sim 10^3\TeV$ in the figure. The second sharp feature near $m_X\sim 10^4\TeV$ corresponds to the transition from the dominance of atmosphere-supplied CHAMPs at low masses to direct-stopping within the pool at higher masses. 

For $m_X \gg 10^{4}\TeV$, only a fraction of the incoming $X$ particles with sufficiently small initial velocities $v_i$ and/or grazing incidence angles stop within the combined atmospheric+pool overburden. In this regime, $N_X^{\rm pool} \propto n_X^{\rm stop}$ with $n_X^{\rm stop}$ given in Eq.~\eqref{eq:nXstop}. For $m_X \ll 10^{11}\,\TeV$ (equivalently, $\bar{L}_{\rm stop} \ll R_\oplus$), this scales as $\propto m_X^{-3}$ for $\bar{L}_{\rm stop} \lesssim \sqrt{R_\oplus h_{\rm pool}}$ and as $\propto m_X^{-1}$ for $\bar{L}_{\rm stop} \gtrsim \sqrt{R_\oplus h_{\rm pool}}$, as found in Eq.~\eqref{eq:nXstopshortLstop}. Finally, for $m_X \gg 10^{11}\,\TeV$ (equivalently, $\bar{L}_{\rm stop} \gg R_\oplus$), it scales as $m_X^{-2}$, as shown in Eq.~\eqref{eq:nXstoplongLstop}.

\subsection{Ancient samples: geological rocks and ice core}

\paragraph{Geological Rocks.---}
Geological rock samples can be extracted from depths of up to $\sim 1\,{\rm km}$, providing enhanced sensitivity to higher-mass CHAMPs, which typically penetrate more deeply before stopping in the Earth (see Fig.~\ref{fig:lenghtscales}). A second advantage is their geological age, which affords many orders of magnitude greater exposure time \cite{Baum:2023cct,Baum:2024eyr,Hirose:2025jht,Graham:2026ivn}. As a result, even modest rock sample volumes can yield sensitivities competitive with those of large, monitored reservoirs (see Fig.~\ref{fig:reach}).

For a robust search using geological samples, the optimal targets are rocks with well-determined ages and depth histories. Standard geological dating techniques, such as radiometric dating, provide reliable age estimates for samples as old as $\sim 4\,{\rm Gyr}$ \cite{wilde2001evidence,bowring1999priscoan}. Geological studies also identify samples with ages $\gtrsim 100\,{\rm Myr}$ that have remained at relatively stable depths $\sim {\rm km}$ \cite{sturrock2024phanerozoic}. Identifying the optimal sample for CHAMP searches is left for future work. In this work, we instead consider two benchmark scenarios for our projected sensitivities, motivated by reported subsidence/uplift rates as well as erosion rates of $\lesssim {\rm km/Myr}$ in slowly evolving geological regions \cite{sturrock2024phanerozoic,wilner2024limits}: 

\begin{enumerate}
    \item a sample that has remained at approximately constant depth throughout a $\sim 10\,{\rm Myr}$ exposure time,
    \item a sample that has migrated in depth over a $\sim {\rm Gyr}$ exposure time.
\end{enumerate}

For the first case, in which the static sample is assumed to remain at a fixed depth, the number of CHAMPs stopped in the sample is given by
\begin{align}
    N_X^{\rm static}=V_{\rm rock}\,\frac{\int_{d_{\rm rock}}^{d_{\rm rock}+L_{\rm rock}}dz\,n_X^{\rm stop}}{L_{\rm rock}} ~,
\end{align}
where $d_{\rm rock}$ denotes the extraction depth, and $L_{\rm rock}$ is the vertical extent (thickness) of the rock sample.

For the second case, in which the sample migrates over the relevant exposure time, we account for geological smearing. We assume that $X$ particles that stop within the range $z \in [0, L_{\rm geo}]$ are uniformly redistributed over this domain over the last Gyr. As a benchmark, we conservatively take $L_{\rm geo}\sim L_{\rm litho} = 100\,{\rm km}$ (in principle $L_{\rm geo}$ could be much smaller than $L_{\rm litho}$, corresponding to less smearing). The number of CHAMPs contained within the fiducial rock sample volume $V_{\rm rock}$ is given by
\begin{align}
    N_{X}^{\rm geo}=V_{\rm rock}\,\frac{\int_{L_{\rm atm}}^{L_{\rm geo}}dz\,n_X^{\rm stop}}{L_{\rm geo}} ~.
\end{align}
In our projections for the $\sim\text{Gyr}$ exposure case, we conservatively adopt the smaller number of CHAMPs obtained from either the direct stopping estimate or the geological-smearing scenario:
\begin{align}
    N_X^{\rm rock}=\text{min}\left(N_X^{\rm static},N_X^{\rm geo}\right) ~.
\end{align}

In Fig.~\ref{fig:reach}, we show the projected sensitivities for geological rocks in both the static (last $\sim 10\text{ Myr}$ exposure) and geologically smeared ($\sim \text{Gyr}$ exposure) cases. We assume a fiducial volume of $1\text{ m}^3$ for the rock sample and show the sensitivities for two different depths, 10 m and 1 km, corresponding to surface and deep rocks, respectively. 

\paragraph{Ancient Ice Core.---}

Ice cores are vertical columns of ice drilled from glaciers in perpetually cold regions such as Antarctica and Greenland. In terms of experimental practicality and projected reach, ice cores are in an intermediate position between monitored water/ice reservoirs and geological rock samples. They offer the dual advantage of being relatively straightforward to enrich and naturally old, though generally not as old as geological rock formations.

The age of the ice varies with depth along the core: deeper layers are older, having formed earlier during the sequential accumulation of snow and its compaction into ice (see Appendix~\ref{app:icecore} for details of a calibrated age–depth relation for ice cores). We compute the total number of $X$ particles by integrating over both the present-day depth and the time-dependent exposure history of each layer,\footnote{If there is a thin layer of water on top the ice core at any given time in the past, it could also catch some airborne CHAMPS, corresponding to rock-equivalent depth $z\leq L_{\rm atm}$. We conservatively neglect this contribution.} 

\begin{align}
    N_X^{\text{core}}=A_{\rm core}\int_{L_{\rm atm}}^{z_{\rm max}} dz\int_0^{t_{\rm age}(z)} dt\, \frac{\partial}{\partial\Delta t} n_X^{\rm stop}[z'(t,z),\Delta t] ~,
\end{align}
where $n_X^{\rm stop}$ is given in Eq.~\eqref{eq:nXstop}, $t_{\rm age}(z)\approx t_{\rm *}\left[z/z_*+(1/20)\left(z/z_*\right)^{20}\right]$ is the age-depth relation fitted from Ref.~\cite{IceCore2023} (see Appendix~\ref{app:icecore}), $t_*$ and $z_*$ are fitting parameters to be specified below, $z_{\rm max}$ denotes the maximum depth of the ice core, $A_{\rm core}$ is its cross-sectional area, $z'(t,z)$ denotes the depth at an earlier time along the ice-core evolution, and $z$ is the present-day depth corresponding to that layer.\footnote{The evolving surface at the bore site implies a time-dependent depth for a given ice element over ${\rm Myr}$ timescales. For simplicity, we approximate the depth evolution of each sample element as linear in time, $z'(t,z)=z\times t/t_{\rm age}(z)$, corresponding to steady accumulation of ice up to its present depth.}

Thus far, ice cores up to a few km in depth and a few Myr in age have been successfully retrieved. For example, the South Pole Ice Core (SPICEcore) project~\cite{johnson2021drilling} recovered a 126 mm diameter, 1751 m deep ice core from the South Pole in 2015. The drilling campaign required 98 active days, corresponding to an effective drilling rate of approximately 25 m/day. For our projected reach, we assume benchmark ice cores with cross-sectional area $A_{\rm core} = 10^{-2}\,{\rm m}^2$, $z_* = 2350\,{\rm m}$, $t_* = 2.5 \times 10^5\,{\rm yr}$, and maximum depth $z_{\rm max}=3\text{ km}$, based on the age-depth relation of the ice core in Ref.~\cite{IceCore2023}; see Appendix.~\ref{app:icecore}.

\section{Detection}
\label{sec:detection}

\subsection{Sample processing}
Large-volume samples collected for a CHAMP search must be {\it enriched} to a high concentration of $X$ particles so that they can be analyzed using a high-precision mass spectrometer. A typical throughput for such instruments is $\sim 10$ microliters per run \cite{Smith:1982qu}. Therefore, the objective of sample processing is to reduce the initial sample volume to a scale manageable by mass spectrometers while retaining the captured CHAMPs within the initial sample.

For liquid samples\footnote{Solid samples identified in this work (rock or ice) must first be melted before the enrichment process begins; see Appendix~\ref{app:enrichment} for details.}, we can employ a {\it gravitational enrichment} strategy. The key idea is that, after reaching the equilibrium distribution (see Appendix~\ref{app:eqtime}), heavy CHAMP particles settle to the bottom of the liquid container within a few Boltzmann height. The enrichment can then be performed iteratively by retaining only the bottom fraction of the liquid and moving it to a thinner container of the same height as the original and then repeating this process multiple times.

The main factors determining the efficiency of gravitational enrichment are the Boltzmann height [Eq.~\eqref{eq:hBoltz}] and the equilibrium timescale (Fig.~\ref{fig:timescale}). Shorter Boltzmann heights and faster equilibration times both increase the enrichment efficiency for higher masses. A concrete implementation of this procedure is discussed in Appendix~\ref{app:enrichment}, where we show that enrichment factors of $\sim 10^{13}$ can be achieved within a year-scale campaign for $m_X \gtrsim 10^{3}\,{\rm TeV}$.\footnote{Note that experimental works performed several decades ago \cite{Smith:1982qu,Pichard:1987ub} achieved enrichment factors of order $\sim 10^{12}$. We expect comparable or better enrichment factors to be achievable with relative ease using modern technology, and even more so for heavier CHAMPs amenable to gravitational enrichment.} With this level of enrichment, an initial water sample volume of $\sim 10^7\,{\rm m}^3$ can be reduced to the milliliter scale (corresponding to gram-scale sample masses). If needed, centrifugation can be employed to further enhance the CHAMP concentration at a significantly faster rate \cite{melting_LHC_2025}.

At the high-mass end of the parameter space, CHAMPs are sufficiently heavy that, during their downward drift, they can gain enough kinetic energy over a typical container height to penetrate the solid layer at the bottom of the container. This occurs for $m_X \gtrsim 2\times 10^{7}\,{\rm TeV}$, as shown in Appendix~\ref{app:eqtime}, where we also demonstrate that such particles stop within a shallow, typically microscopic depth of metals. To accommodate this mass range, we propose lining the inner surfaces of containers used during sample processing with a thin layer of low–melting-point metal (e.g., tin) to capture these particles. The metal lining can subsequently be collected and processed with techniques optimized for heavier CHAMPs.

For $m_X \lesssim 10^3\,{\rm TeV}$, gravitational enrichment becomes inefficient because the Boltzmann height is large ($h_{\rm Boltz} \gtrsim {\rm m}$) and the equilibrium timescale is long ($t_{\rm eq}\gtrsim {\rm yr})$. In this regime, we restrict the initial sample volume to smaller sizes ($\sim 10^5\,{\rm m}^3$), which can be processed directly using centrifugal enrichment, as detailed in Ref.~\cite{melting_LHC_2025}.

For ancient rock samples, the long exposure time compensates for the smaller initial sample size. Although additional steps are required to melt the sample into a liquid phase (see Appendix~\ref{app:enrichment}), the required starting volume, and thus the necessary enrichment factor, can be significantly less stringent. For example, a $\sim 100\,{\rm Myr}$-old sample requires an enrichment factor smaller by $\sim 10^8$ to achieve the same sensitivity to CHAMP abundance as a sample monitored in real time for one year, simply because one can begin with a proportionally smaller initial volume.

\subsection{CHAMP identification}
The outcome of the sample processing and enrichment in our proposed strategy is a gram-scale sample (corresponding to a volume of order ${\rm cm}^3\sim{\rm milliliter}$) that contains essentially all the CHAMP particles initially trapped in the much larger starting volume. The next step is to identify CHAMP particles in this sample.

{\it Anomalously heavy sample.---} For sufficiently heavy CHAMP particles, their large mass concentration manifests itself as an anomalously large total mass of the enriched sample, detectable by simple weighing \cite{Smith:1982qu}. This provides a quick, low-cost, and non-destructive probe for the presence of CHAMP particles. While a positive signal in such a measurement would not by itself constitute a definitive discovery, it serves as a preliminary test and consistency check before proceeding to high-precision analyses. With commercial precision balances, mass sensitivities of order $\sim 0.1\,\mu\mathrm{g}$ can be achieved for gram-scale samples \cite{Mettler_Toledo,Sartorius}. Therefore, if the final enriched sample contains $N_X$ CHAMPs of mass $m_X$, an anomalous excess mass (compared to pre-enriched sample mass of the same volume) is detectable if $N_Xm_X\gtrsim0.1\,{\rm \mu g}$. This corresponds to a sensitivity of
\begin{equation}
    N_X \gtrsim 5\times10^3 \left(\frac{m_X}{10^{10}\,{\rm TeV}}\right)^{-1} ~,
\end{equation}
with the caveat that this measurement is not a conclusive signature of CHAMPs, since in principle other heavy compounds or molecules could also be present in the final sample, potentially leading to ambiguity.

{\it Mass spectrometry.---} A robust identification of CHAMPs can be achieved using mature mass spectrometry techniques, which typically proceed through the following steps: vaporization of the sample, ionization of the resulting atomic or molecular vapor, accelerating the resulting ions in a background electric field, followed by suppression of standard model heavy atomic backgrounds via magnetic deflection, and measurement of the charge-to-mass ratio of the particles that pass the magnetic-deflection filter. 

Mass spectrometry is a well-established technology with a wide range of optimization techniques and modifications developed for different applications. It has previously been used in searches for CHAMP particles \cite{Smith:1982qu,Yamagata:1993jq,Middleton:1979zz}, and some of us have recently studied and reviewed the use of mass spectroscopy for CHAMP searches in Ref.~\cite{melting_LHC_2025}. In the present work, similar methods can also be applied. Below, we discuss several aspects of applying mass spectrometry in the mass range of interest that require further consideration and care.

A major challenge, as noted in previous works on CHAMP searches, is the background from heavy multi-atom molecules. For example, proteins and viruses can reach masses as large as $\sim 10^{6}\,{\rm TeV}$, in which case they constitute a potential background in the lower half of the mass range considered in this work. Various techniques have been developed to break molecular bonds and mitigate such backgrounds by fragmenting large molecules into smaller constituents. One class is collision-induced dissociation, in which ions are passed through a thin film or, more commonly, a low-pressure inert gas medium, where collisions excite vibrational modes of the molecules and ultimately lead to bond breaking. A second class is to actively break the molecules, such as accelerator mass spectrometry (AMS)~(reviewed in Refs.~\cite{Burdin:2014xma,Perl:2001xi,strangelets_search_review_2005}), which employs an $\sim\text{MeV}$ tandem accelerator, or ultraviolet photodissociation, where high-energy photons are used to break molecular bonds. A third class is hard ionization~\cite{Ionization_MassSpec_Guide}, which breaks molecular bonds at the ionization step, e.g., electron ionization
(EI) uses a current whose de Broglie wavelength matches molecular bond lengths~\cite{70eV_bond_break}. One or a combination of these methods may be employed in a final experimental design to maximally suppress this background.

The velocity acquired by a CHAMP in a typical mass-spectrometer bias field of $E_{\rm bias}\sim{\rm kV}$ is $\sqrt{E_{\rm bias}/m_X}\sim 20\,{\rm cm/s} ~(m_X/10^{10}\,{\rm TeV})^{-1/2}$. Heavier CHAMPs therefore remain extremely slow even after acceleration. In this regime, conventional time-of-flight mass spectrometry (TOF-MS), with microchannel plates as ion detectors, is ineffective since impact ionization is strongly suppressed at such low velocities.

Charge detection mass spectrometry (CDMS) is a more suitable alternative technology for higher-mass ions. In CDMS, the accelerated ion passes through a conducting cylinder, inducing an image charge that can be precisely amplified and measured. The signal amplitude determines the charge, while the pulse duration (transit time) determines the mass. Commercial CDMS instruments can identify ions with masses up to $\sim 10^{5}\,{\rm TeV}$ \cite{CDMS}, and laboratory implementations have demonstrated sensitivity up to $\sim 10^{8}\,{\rm TeV}$ \cite{hrabovsky2024charge,jarrold2021applications}. In principle, since CDMS employs charge-sensitive amplification circuits, which do not directly depend on velocity/mass, even heavier CHAMPs can be detected. 

In current biotechnology applications, extending CDMS sensitivity to higher masses faces practical challenges, such as the large physical size of protein complexes (which is not relevant for the point-like CHAMP particles of interest here) and gravitational free fall during the measurement time. The latter effect is also present in CHAMP searches, but it can be mitigated in a custom-designed setup that operates in the vertical direction, aligned with gravity.\footnote{There are further opportunities to improve the detection efficiency. For example, a transition-edge sensor (TES) placed at the end of the CDMS stage could measure the total kinetic energy of the CHAMP. In such a multi-modal setup, one could independently reconstruct charge, mass, and kinetic energy, enabling consistency checks. Furthermore, an array of spatially-separated superconducting quantum interference devices (SQUIDs) can detect the passage of a charged particle and directly measure its charge \cite{Gao:2025ryi}. Other technologies, such as ion traps \cite{Schmid:2017eyj,Budker:2021quh,Carney:2021irt}, optically levitated sensors \cite{Moore:2014yba}, or trapped-ion crystals, can be employed in a similar manner, functioning as sensitive passing-charge detectors and timing devices.} Another major advantage of nondestructive methods like CDMS is that particles can be captured upon detection and further tested, arbitrarily increasing the statistical significance~\cite{melting_LHC_2025}.

\section{Discussion and conclusion}
\label{sec:conclusion}

We propose terrestrial search strategies for galactic, virialized, CHArged Massive Particle (CHAMP) relics that stop and accumulate in the Earth, with sensitivity to a CHAMP population comprising even a tiny fraction $f_X$ of dark matter. As CHAMPs traverse the atmosphere and Earth matter, they lose energy and eventually come to rest in terrestrial reservoirs. We compute the distribution of first-stopping points as a function of rock-equivalent depth below the Earth's interface, and use it to identify target materials and experimental protocols that maximize reach and mimize systematics.

A major challenge for terrestrial searches is the \textit{a priori} unknown spatial distribution of stopped CHAMPs in natural environments. We address this by estimating the distribution of CHAMP stopping points and focusing on targets with controlled or suppressed transport histories: monitored water pools, monitored natural ice, geological rocks, and ancient ice cores. Monitored water pool with surface exposed to air has the additional advantage that it can capture CHAMPs that stop in the atmosphere, neutralize, and sediment under gravity.

A shown in Fig.~\ref{fig:reach}, our projected sensitivity spans many orders of magnitude in CHAMP galactic abundance beyond existing limits, for masses from 1 TeV to $10^{9}$~TeV. A practical experimental program would begin with modest sample volumes to establish a complete search pipeline. Once validated, it can be scaled up to achieve the ultimate projected reach and discovery potential. Two illustrative pathfinder strategies are as follows: In the low-mass regime ($m_X \lesssim 10\,{\rm TeV}$), our full projections using $V_{\rm water} \sim 10^5\, {\rm m}^3$ and an exposure time $t_{\rm exp} \sim 10\,{\rm yr}$ reach a sensitivity of $f_X \sim 10^{-20}\text{--}10^{-18}$. Scaling down to a pathfinder experiment with $V_{\rm water} \sim (10\,{\rm cm})^3$ (a volume for which centrifugal enrichment is straightforward even with a single centrifuge) and $t_{\rm exp} \sim \mathrm{month}$ would yield sensitivity at the level of $f_X \sim 10^{-11}\text{--}10^{-9}$. For higher masses ($m_X \sim 10^4\,{\rm TeV}$), where gravitational enrichment is efficient, a pathfinder experiment with $V_{\rm water} \sim {\rm m}^3$ and $t_{\rm exp} \sim {\rm month}$ could also achieve a sensitivity of $f_X \sim 5\times 10^{-12}\text{--}5\times 10^{-10}$. Each of these pathfinder experiments already probes a significant new parameter space beyond existing constraints.

Particle track detectors that rely on real-time detection of energy deposition or on imaging of permanent lattice damage in solids \cite{Price:1984fk,Price:1986ky,Snowden-Ifft:1995zgn,Ebadi:2021cte,Baum:2023cct,Baum:2024eyr,Hirose:2025jht,Graham:2026ivn} are less suited for detecting the CHAMPs we consider for two reasons. First, CHAMPs may not induce detectable energy depositions or lattice damages. Track formation requires hard nuclear collisions that displace atoms from lattice sites \cite{Rajendran:2017ynw,Marshall:2020azl,Marshall:2021xiu,Ebadi:2022axg,Ebadi:2021cte}, whereas our CHAMPs lose energy primarily through soft electromagnetic interactions, transferring only tiny amounts of energy per electron which leads to hardly any defects. Second, even if these energy depositions or lattice damages were detectable, achieving competitive sensitivity would require processing and imaging extremely large volumes of sample, which is highly impractical with existing track-based approaches. Third, our method could in principle be background-free, because there is no naturally occurring ultra-heavy atoms.

While we focus on singly charged CHAMPs $X^\pm$ for concreteness, much of the discussion is broadly applicable to other long-lived particles, including multiply charged particles, millicharged particles, color-charged particles, and magnetic monopoles, given appropriate modifications to the capture and identification steps \cite{Burdin:2014xma}. There might even be practical application if some such particles were discovered: it has been argued that doubly-negative charged particles $X^{--}$ could be used for cold fusion~\cite{Akhmedov:2021qmr}.

Our stopping-depth-distribution calculation techniques may also be useful beyond the specific searches proposed here. They can be incorporated into sensitivity estimates for solid-state track-based experiments and related overburden-limited probes, e.g., for macroscopic dark matter \cite{Price:1984fk,Price:1986ky,Snowden-Ifft:1995zgn,Ebadi:2021cte,Baum:2023cct,Baum:2024eyr,Hirose:2025jht, Rajendran:2017ynw,Marshall:2020azl,Marshall:2021xiu,Ebadi:2022axg,Kaplan:2024dsn,Fedderke:2024hfy}.

\section*{Acknowledgements}
We thank Michael Fedderke for useful discussion.

This work was supported by the U.S.~Department of Energy~(DOE), Office of Science, National Quantum Information Science Research Centers, Superconducting Quantum Materials and Systems Center~(SQMS) under Contract No.~DE-AC02-07CH11359.

R.E. is supported by the John Templeton Foundation Award No. 63595, the University of Delaware Research Foundation, and NSF Grant No. PHY-2515007. The work of R.E.
was also supported by the Grant 63034 from the John
Templeton Foundation. R.E. gratefully acknowledges the
Pacific Postdoctoral Program at the Dark Universe Science Center, University of Washington, where part of
this work was carried out. The Pacific Postdoctoral Program is supported by a grant from the Simons Foundation (SFI-MPS-T-Institutes-00012000, ML).

S.R.~ is supported in part by the U.S.~National Science Foundation~(NSF) under Grant No.~PHY-1818899, 
the Simons Investigator Grant No.~827042.

This work was performed in part at the Aspen Center for Physics, which is supported by National Science Foundation grant PHY-2210452.

S.~W. is supported in part by the U.S. Department of Energy Office of Science under Award Number DE-SC0024375; the Gordon and Betty Moore Foundation through Grant GBMF13898 to the University of Washington; and the J.~J., L.~P., and A.~J. Smortchevsky Fellowship. S.~W. is grateful to the Mainz Institute for Theoretical Physics (MITP) of the Cluster of Excellence PRISMA+ (Project ID 390831469), for its hospitality and its partial support during the completion of this work. This research was supported in part by Perimeter Institute for Theoretical
Physics. Research at Perimeter Institute is supported by the Government of Canada through the Department of Innovation, Science and Economic Development and by the Province of Ontario through the Ministry of Colleges
and Universities.

This work was supported in part by NSF Grant No. PHY-2310429, Simons Investigator Award No. 824870, and the John Templeton Foundation Award No. 63595.

\appendix
\section{Distribution of stopping depths}
\label{app:stoppingdist}
In this appendix, we present details of the computation of the CHAMP stopping distribution. We assume straight-line trajectories, neglect gravitational effects (by only including incoming particles with speeds above the Earth's escape speed), and ignore deflections due to solar wind and the Earth’s magnetic field, which are negligible in the mass range of interest due to the large particle momenta \cite{Cholis:2015gna,Potgieter:2013mcc}.

\subsection{Stopping power}
We are primarily interested in $X$ particles with velocities $v \lesssim 10^{-3}$. These particles are slower than typical bound electrons in solids, which have velocities $v\lesssim \alpha_{\rm em}= 7\times 10^{-3}$. In this regime, the commonly used Bethe–Bloch stopping formula is not applicable. Instead, we adopt the Lindhard–Scharff stopping power, assuming a projectile atomic number $Z_1 = 1$ and a target atomic number $Z_2 = \mathcal{O}(10)$ \cite{ParticleDataGroup:2024cfk,Burdin:2014xma,Lindhard:1961zz}:\footnote{Eq.~\eqref{eq:dEdx} is obtained by extrapolating linearly in $v$ the muon stopping range in copper (PDG's Fig.~34.1) at $v=10^{-3}$ to lower speeds \cite{ParticleDataGroup:2024cfk}.}
\begin{align}\label{eq:dEdx}
    \left|\frac{dE}{dx}\right|\approx 10~\text{GeV}/\text{m}\times \left(\frac{v}{10^{-3}}\right)\left(\frac{\rho_{\rm matter}}{2.5~\text{g}/\text{cm}^3}\right) ~.
\end{align}
The stopping length for an $X$ particle with initial velocity $v_i \lesssim 10^{-3}$ is
\begin{align}
    L_{\rm stop}(v_i)&= \int_{0}^{v_i}dv\frac{m_Xv}{|dE/dx|}\nonumber\\
    &\approx 0.1~\text{mm}\left(\frac{v_i}{10^{-3}}\right)\left(\frac{m_X}{1~\TeV}\right)\left(\frac{\rho_{\rm matter}}{2.5~\text{g}/\text{cm}^3}\right)^{-1} ~.
\end{align}

Note that for the mass range of interest $m_X\lesssim 10^{14}\TeV$, the deceleration due to this stopping power in Eq.~\eqref{eq:dEdx} is $\gtrsim 10\,{\rm m/s^2}$. This is greater than the gravitational acceleration. This justifies assuming that the motion of the $X$ particle in terrestrial matter is dominantly set by its initial velocity and the stopping power.

$X^+$ captures an electron while $X^-$ captures an ion by binding to a nucleus (together with the electrons bound to it). Thus, once neutralized, their cross section could in principle be different. The stopping power due to hard scatterings with nuclei via strong interactions can be estimated as $dE/dx\sim (n_{N}\sigma_{\rm strong}v)(m_Nv)\sim \rho_{\rm matter}\sigma_{\rm strong}v^2\sim 3\times 10^{-6}\GeV/\text{m} \times (v/10^{-3})^2$, where $m_N\sim\GeV$, $\sigma_{\rm strong}\sim \text{fm}^2$, and $\rho_{\rm matter}=5~\text{g}/\text{cm}^3$. Therefore, the stopping power due to the strong interaction is negligible compared to that of electromagnetic interactions. This result also suggests that our stopping-distribution results for $X^+$ also apply to $X^-$ particles (whose bound state with nuclei experiences nuclear scattering), as their electromagnetic interactions and thus stopping power are similar.

\begin{figure}[t!]
    \centering
    \includegraphics[width=\linewidth]{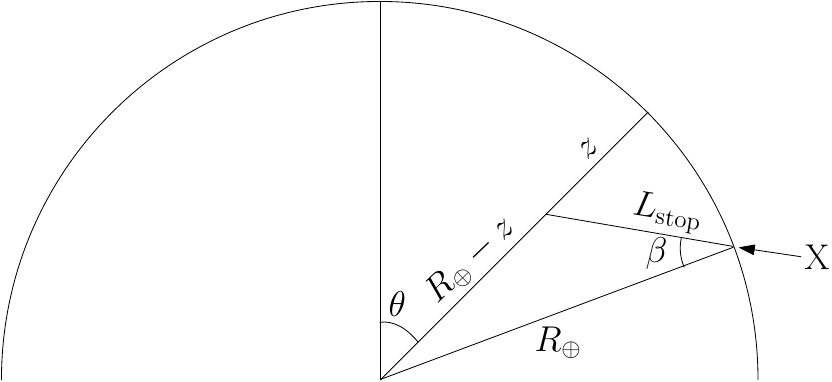}
    \caption{CHAMP stopping geometry. The $X$ particle enters the Earth and stops after traveling a distance $L_{\rm stop}$, corresponding to a stopping rock-equivalent depth $z$ below the Earth's surface (measured from the top of the troposphere). $0<\theta<\pi$ is the polar angle and $-\pi/2<\beta<\pi/2$ is the entry angle.}
    \label{fig:geometry}
\end{figure}

\subsection{Dirac-delta velocity distribution}
Throughout, we assume for simplicity that the initial velocity distribution of $X$ is isotropic. We begin by considering a monochromatic initial-speed distribution 
\begin{align}
    f(v_i)=\delta(v_i-v_{\rm DM}) ~,
\end{align}
and a corresponding stopping length $L_{\rm stop}=L_{\rm stop}(v_{\rm DM})$. In a given time interval $\Delta t$, the number of particles stopping in a volume spanning a differential solid angle $\Delta \Omega_\theta$ and depth range $(z,z+\Delta z)$ is given by
\begin{align}
    n_X^{\rm stop}(R_\oplus-z)^2\Delta\Omega_\theta\Delta z=\Phi_X\Delta t   \frac{\Delta\Omega_\beta}{4\pi}(R_\oplus^2\Delta\Omega_\theta\cos\beta) ~,
\end{align}
where the depth $z$ and the angles $\theta$ and $\beta$ are defined in Fig.~\ref{fig:geometry}. Here, the factor $\Delta\Omega_\beta/4\pi$ captures the fraction of incoming CHAMPs whose directions are compatible with stopping at depth $z$, and the factor $R_\oplus^2\Delta\Omega_\theta\cos\beta$ represents the effective Earth surface area perpendicular to the incoming particles. The solid angle for $\theta$ is defined as $\Omega_{\theta}=2\pi(1-\cos\theta)$, and similarly for $\beta$. The accumulated number density of $X$ over a time interval $\Delta t$ can thus be found as
\begin{align}
    n_X^{\rm stop}&=\frac{\Phi_X\Delta t}{2}\frac{R_\oplus^2}{(R_\oplus-z)^2}\cos\beta\frac{d\cos \beta}{dz}\nonumber\\
    &=\frac{\Phi_X\Delta t}{4L_{\rm stop}}\frac{2R_\oplus z+L_{\rm stop}^2-z^2}{(R_\oplus-z)L_{\rm stop}}\Theta(L_{\rm stop}-z) ~,
\end{align}
where in the last step we make use of the cosine rule,
\begin{align}
\cos\beta=\frac{R_\oplus^2+L_{\rm stop}^2-(R_\oplus-z)^2}{2R_\oplus L_{\rm stop}} ~,
\end{align}
where $L_{\rm stop}/2R_\oplus\leq\cos\beta\leq1$, and $\cos\beta$ is a monotonically increasing function of $z$. The lower and upper bounds of $\cos\beta$ correspond to $z=0$ and $z=L_{\rm stop}$, respectively. We are mainly interested in practically accessible depths that are less than a few km from Earth's surface; therefore, $z\ll R_\oplus$, for which the $n_X^{\rm stop}$ simplifies to
\begin{align}
n_X^{\text{stop}}=\frac{\Phi_X\Delta t}{4R_\oplus}\left(1+\frac{2R_\oplus z}{L_{\rm stop}^2}\right)\Theta(L_{\rm stop}-z)~.
\label{eq:nXDiracDelta}
\end{align}
When $L_{\rm stop}^2/R_\oplus\ll z\ll R_\oplus$, the flat-Earth approximation holds.

\subsection{Maxwell-Boltzmann velocity distribution}
For sufficiently large $m_X$, typical particles with $v_i \sim v_{\rm DM}$ have $L_{\rm stop} \gg  2R_{\oplus}$, implying that they traverse the Earth without stopping. In this regime, only particles in the low-velocity tail of the distribution can be captured in the Earth. This motivates considering the full (assumed Maxwell-Boltzmann) velocity distribution rather than focusing only on typical particles with $v_i\sim 10^{-3}$. Note that the Maxwell–Boltzmann distribution exhibits a power-law behavior at low velocities, in contrast to the much stronger exponential suppression at high velocities.

Substituting $\Phi_X$ with $\int dv_i\, f(v_i) \Phi_X$ and $L_{\rm stop}$ with $ L_{\rm stop}(v_i)$ in Eq.~\eqref{eq:nXDiracDelta}, and integrating over the initial speeds $v_i$, we have
\begin{align}
    n_X^{\rm stop}=&\frac{\Phi_X\Delta t}{4R_\oplus}\int_0^{\infty}dv_i\,f(v_i)\left(1+\frac{2R_\oplus z}{L_{\rm stop}^2(v_i)}\right)\nonumber\\
    &\times \Theta[L_{\rm stop}(v_i)-z]\Theta[2R_\oplus-L_{\rm stop}(v_i)]\nonumber\\
    =&\frac{\Phi_X\Delta t}{4R_\oplus}\int_{v_{\rm min}}^{v_{\rm max}}dv_i\,f(v_i)\left(1+\frac{2R_\oplus z}{L_{\rm stop}^2(v_i)}\right)\nonumber\\
   =&\frac{\Phi_X\Delta t}{4R_\oplus}\left\{\left[\text{Erf}\left(\frac{v_{\rm max}}{\sqrt{2}\sigma_v}\right)-\text{Erf}\left(\frac{v_{\rm min}}{\sqrt{2}\sigma_v}\right)\right]\left(1+\frac{4R_\oplus z}{\bar{L}_{\rm stop}^2}\right)\right.\nonumber\\
    &\left.+\frac{2}{\sqrt{\pi}}\left(\frac{v_{\rm max}}{\sqrt{2}\sigma_v}e^{-\frac{v_{\rm max}^2}{2\sigma_v^2}}-\frac{v_{\rm min}}{\sqrt{2}\sigma_v}e^{-\frac{v_{\rm min}^2}{2\sigma_v^2}}\right)\right\} ~,
\end{align}
where we assume $z\ll R_\oplus$ [as is assumed in Eq.~\eqref{eq:nXDiracDelta} as well], $\sigma_v=10^{-3}$, and we define
\begin{align}
    v_{\rm min}&\equiv \text{max}\left(\sqrt{2}\sigma_v\frac{z}{\bar{L}_{\rm stop}},v_{\rm esc,\oplus}\right) ~,\\
    v_{\rm max}&\equiv \sqrt{2}\sigma_v\frac{2R_\oplus}{\bar{L}_{\rm stop}} ~,\\
    \bar{L}_{\rm stop}&\equiv L_{\rm stop}(v_i=\sqrt{2}\sigma_v) ~.
\end{align}
As mentioned in the main text, we neglect particles with incoming speeds smaller than the Earth’s escape velocity $v_{\rm esc,\oplus} = 4\times 10^{-5}$, since the distribution of stopping points for such particles is significantly affected by Earth’s gravity, which complicates their dynamics. This requirement is reflected in the expression for $v_{\rm min}$ above.

We further impose $v_{\rm min} \lesssim v_{\rm max}$, which implies $\bar{L}_{\rm stop} \lesssim 2R_\oplus \times (\sqrt{2}\sigma_v / v_{\rm esc,\oplus})$, and thus
\begin{align}
    m_X\lesssim 3\times 10^{12}\TeV\,.
\end{align}
For masses larger than this, no particles can stop in the Earth in our approximation since even the lowest velocity particles entering at the Earth's escape velocity will not stop in one straight passing/orbit.

\section{Equilibration timescale}

\label{app:eqtime}

\begin{figure}
    \centering
    \includegraphics[width=\linewidth]{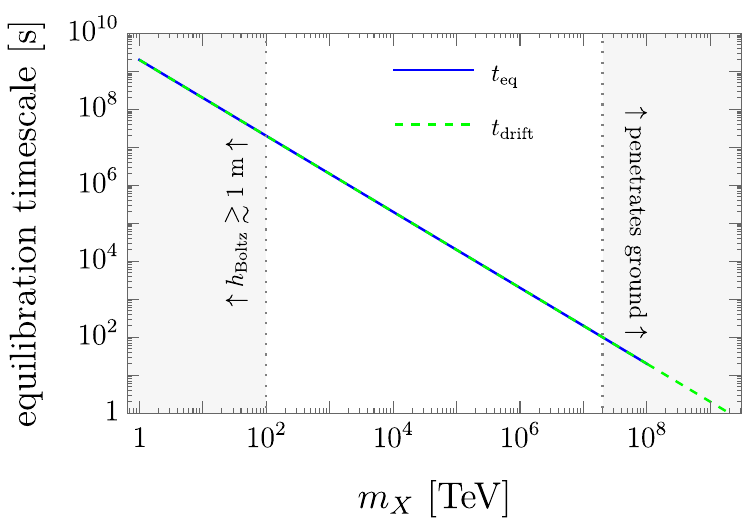}
    \caption{Boltzmann equilibration timescale in water with height $H=10~\text{m}$, relevant for CHAMP accumulation in a stopping target and gravitational enrichment.}
    \label{fig:timescale}
\end{figure}

After the CHAMPs $X^\pm$ have slowed down to near stopping due to Lindhard-Scharff energy loss in a fluid, they will capture an electron or an ion $A^+$ and form neutral bound states, $X^+e^-$ or $X^-A^+$, which do not experience the Lindhard-Scharff stopping. The $X$ particles will then redistribute themselves toward the Boltzmann distribution via a dynamics governed by gravitational acceleration and repeated elastic collisions with atoms/molecules of the medium. In this appendix, we estimate the timescale to achieve the Boltzmann distribution, which is relevant for CHAMP accumulation in a stopping target and gravitational/centrifugal enrichment considerations. We will assume, for concreteness, that the medium has a temperature $T=300~\text{K}$, a density $\rho_A^{(\text{water})}=1~\text{g/cm}^3$ ($\rho_A^{(\text{air})}=10^{-3}~\text{g/cm}^3$) for water (air), and nuclear mass $m_A=20\GeV$, although our results can be trivially generalized to other parameters. The resulting equilibration timescale is summarized in Fig.~\ref{fig:timescale}.

\subsection{Biased random walk regime {\boldmath ($m_X\lesssim 4\times 10^{6}\TeV$)}}

In this regime the heavy neutral bound state undergoes a random walk with a small downward bias due to gravity, well captured by the biased-diffusion (Drude) model.

For a room-temperature ($T=300~\text{K}$) medium, the thermal speed of a typical atom is $v_{A,\rm th}\sim \sqrt{T/m_A}\sim 1\times 10^{-6}$. In all cases of our interest, the neutralized heavy particle moves much more slowly than atoms, $v_X\ll v_{A,\rm th}$, so the collision rate of the heavy particle with atoms is set by $v_{A,\rm th}$ rather than $v_X$. We model the collisions as billiard-ball-like scattering with geometric cross section $\sim a^2$, where $a\equiv \AA$. Over a time interval $\Delta t$, the number of collisions is $\sim (\rho_A/m_A)\times a^2\times v_{A,\rm th}\times \Delta t$. Each collision imparts a momentum $\Delta p_{X,1}\sim m_A v_{A,\rm th}$ to the heavy particle, in a random direction. Thus the heavy particle random walks in the momentum space as $\Delta p_X\sim \Delta p_{X,1}\sqrt{(\rho_A/m_A) a^2 v_{A,\rm th}\Delta t}$, which implies a momentum-transfer mean free time of $\tau_{\rm tr}\sim m_X/(\rho_A a^2v_{A,\rm th})$, defined as the time for the expected momentum change to accumulate to $\Delta p_X\sim m_Xv_{X,\rm th}$ with $v_{X,\rm th}\sim \sqrt{T/m_X}$. Gravity then induces a downward drift velocity $v_{\rm drift}\sim g\tau_{\rm tr}$:
\begin{align}\label{eq:vdrift}
    v_{\rm drift}\sim\frac{m_X g}{\rho_A a^2v_{\rm A,th}}\sim 2\times 10^{-10}\left(\frac{m_X}{10^{7}~\TeV}\right)\left(\frac{\rho_A}{1~\text{g/cm}^3}\right)^{-1} ~.
\end{align}

The Drude picture requires that the drift be a small perturbation to thermal motion, $v_{\rm drift}\lesssim v_{X,\rm th}\sim \sqrt{T/m_X}$, which translates to $m_X\lesssim (\rho_A^2a^4T^2/m_Ag^2)^{1/3}\sim  4\times 10^6\TeV\,[\rho_A/(1~\text{g/cm}^3)]^{2/3}$, i.e., $m_X\lesssim 4\times 10^6\TeV$ ($m_X\lesssim 4\times 10^4\TeV$) in water (air).

In the Drude regime, the diffusion time across a scale height $H$ is $t_{\rm diff}\sim H^2/(v_{X,\text{th}}^2\tau_{\rm tr})\sim \rho_Aa^2H^2/\sqrt{Tm_A}$, and the timescale to drift vertically by $H$ is $t_{\rm drift}\sim H/v_{\rm drift}\sim \rho_A a^2H\sqrt{T/m_A}/(m_X g)$. For $m_A\sim 20\GeV$, $T=300~\text{K}$, and $g=10~\text{m/s}^2$, we have
\begin{align}
    t_{\rm diff}&\sim 3~\text{yr}\left(\frac{H}{10~\text{m}}\right)^2\left(\frac{\rho_A}{1~\text{g/cm}^3}\right)~,\\
    t_{\rm drift}&\sim 0.6~\text{yr}\left(\frac{m_X}{100~\TeV}\right)^{-1}\left(\frac{H}{10~\text{m}}\right)\left(\frac{\rho_A}{1~\text{g/cm}^3}\right) ~.
    \label{eqn:drifttime}
\end{align}

If $h_{\rm Boltz}\lesssim H$ ($h_{\rm Boltz}\gtrsim H$), then the distribution of $X$ particles needs to compress (expand) vertically to reach the Boltzmann distribution $n_X\propto e^{-h/h_{\rm Boltz}}$ with $h_{\rm Boltz}=T/(m_Xg)$. Thus, the equilibration timescale is set by $t_{\rm drift}$ for $h_{\rm Boltz}\lesssim H$ and by $t_{\rm diff}$ for $h_{\rm Boltz}\gtrsim H$. Incidentally, $h_{\rm Boltz}\lesssim H$ implies $t_{\rm drift}\lesssim  t_{\rm diff}$. The equilibration timescale can thus be summarized as the minimum of the two timescales,
\begin{align}
    t_{\rm eq}=\text{min}\left(t_{\rm drift},t_{\rm diff}\right); \quad (m_X\lesssim 4\times 10^6\TeV)   ~.
\end{align}

Gravitational enrichment is effective only when $h_{\rm Boltz}\ll H=10\text{ m}$ (corresponding to $m_X\gg 10^3\TeV$ for $T=300~\text{K}$). For capture by a water pool of height $H$, it is possible to have $h_{\rm Boltz}\gtrsim H$. However, diffusion in the air above the pool toward an $\sim h_{\rm Boltz}$ spread occurs very rapidly, and so the bottleneck of the process is still the slow diffusion within the water of column height $H$. Consequently, for an accumulation time $\lesssim 10 ~\text{yr}$ particles that stop in a quiet pool of height $H=10~\text{m}$ will largely remain within the pool even when $h_{\rm Boltz}\gg H$.

\subsection{Direct fall with drag regime {\boldmath ($m_X\gtrsim 4\times 10^6\TeV$)}}

For $m_X \gtrsim 4\times 10^6$~TeV, the drift velocity in water $v_{\rm drift}$ becomes faster than the thermal velocity $v_{X,\rm th}$, and the Drude model breaks down. In this regime, a neutral bound state containing an $X$ particle falling with a velocity $v_X$ in a fluid with atomic mass density $\rho_A$  experiences a collisional drag force 
$\mathbf{F}_{\rm drag}\sim (\rho_A/m_A)\times a^2\times \int d^3v_A\,f(v_A)|\mathbf{v}_A-\mathbf{v}_X|[m_A(\mathbf{v}_A-\mathbf{v}_X)]$ in the limit $m_A\ll m_X$. Atoms coming from the front of $X$ have a slightly larger relative speed than those from behind. This leads to asymmetries in the collision rates and the leading $\mathcal{O}(v_A^2)$ term in $F_{\rm drag}$ cancels and the surviving term is suppressed by $\sim v_X/v_{A,\rm th}$. So $F_{\rm drag}\sim (\rho_A/m_A)\times a^2\times v_{\text{A,th}}\times m_Av_{\text{A,th}}\times (v_X/v_{\text{A,th}})$. The fastest $X$ can be accelerated gravitationally after neutralization is $v_X\sim \sqrt{2gH}\sim 10^{-8}\sqrt{H/10~\text{m}}$, so indeed $v_X\ll v_{\rm A,th}\sim \sqrt{T/m_A}\sim 1\times 10^{-6}$.

The vertical equation of motion of the heavy particle is therefore 
\begin{align}
    \frac{dv_X}{dt}\sim g-\frac{\rho_A a^2v_{\rm A,th}}{m_X}v_X ~, 
\end{align}
which admits a terminal velocity $v_{\rm term}$ that is parametrically the same as the drift velocity in the Drude regime $v_{\rm drift}$ as given in Eq.~\eqref{eq:vdrift}, $v_{\rm term}\sim v_{\rm drift}$. An $X$ particle that starts at $v_{X,\rm th}\lesssim v_{\rm term}$ would speed up to the terminal velocity $v_{\rm term}$ by gravity (which initially dominates over the drag) over a distance
\begin{align}
    d_{\rm term}\sim \frac{v_{\rm term}^2}{g}\sim 0.36\text{ mm}\left(\frac{m_X}{10^{7}~\TeV}\right)^2\left(\frac{\rho_A}{1~\text{g/cm}^3}\right)^{-2} ~.
\end{align}
Thus, the terminal velocity $v_{\rm term}$ is reached within a distance  $d_{\rm term}\lesssim 10~\text{m}$ (the fiducial height of an enrichment tank) for 
\begin{align}
    m_X\lesssim 2\times 10^{9}\TeV\quad\text{(reaches }v_{\rm term}\text{ before ground})  ~.  
\end{align}
If the terminal velocity $v_{\rm term}$ is not reached within the available fall height $H$, the impact speed is instead set by free fall, $v_{\rm fall}\sim \sqrt{2gH}\sim 10^{-8}\sqrt{H/10~\text{m}}$. Accounting for both possibilities, the ground-impact velocity is 
\begin{align}\label{eq:vground}
    v_{\rm ground}\sim \text{min}\left(v_{\rm term},  10^{-8}\sqrt{\frac{H}{10~\text{m}}}\right)~,
\end{align}
and the time to reach the ground is
\begin{align}\label{eq:tground}
    t_{\rm ground}\sim \text{max}\left[t_{\rm term}, 1~\text{s}\left(\frac{H}{10~\text{m}}\right)^{-1/2}\right]~.
\end{align}
where $v_{\rm term}\sim v_{\rm drift}$ and $t_{\rm term}\sim H/v_{\rm term}\sim t_{\rm drift}$ (the times it takes for the particle to reach the ground if $v\approx v_{\rm term}$ during most of the particle's descent) are given by Eqs.~\eqref{eq:vdrift} and \eqref{eqn:drifttime}, respectively.

\subsubsection{Penetration criterion}

Upon reaching a solid ground, the neutral bound state may bounce off or plow into it. Consider a heavy bound state impacting the solid with speed $v$. After a time $t$ since impact, the heavy particle would advance a depth $\sim vt$ if it is not stopped earlier. Sound waves propagate radially at the sound speed $c_s\sim 10^{-5}$ (which is $\gg v$ in the cases we consider), and so the radius of the affected region grows as $c_st$. If we assume that the material compresses by this distance $\sim vt$, then the bulk deformation energy stored in this region is
\begin{align}
    KV\left(\frac{\Delta V}{V}\right)^2 \sim K c_s v^2t^3 ~,
\end{align}
where $K\sim 100~\text{GPa}\sim \eV/\AA^3$ is the typical bulk modulus of a solid, $\Delta V\sim vt\times (c_st)^2$, and $V\sim (c_st)^3$. The heavy particle comes to a stop when the above becomes the initial kinetic energy of the particle $\sim m_X v^2$, which occurs after a time
\begin{align}
    t_{\rm stop}\sim \left(\frac{m_X}{Kc_s}\right)^{1/3}\sim 10~\text{ps}\left(\frac{m_X}{10^7\TeV}\right)^{1/3} ~.
\end{align}

The heavy particle will penetrate if the compressed solid cannot exert sufficient force to slow the particle down fast enough.  Thus,
requiring the rate of change of the heavy particle's momentum (which is $\sim m_Xv/t_{\rm stop}$) be greater than the largest force the atom right beneath it can sustain (which is $\sim  \eV/\AA$), and also requiring $m_X v^2/2\gtrsim \eV$ (or, equivalently, $vt_{\rm stop}\gtrsim \AA$), leads to the following condition for penetration:
\begin{align}
    v\gtrsim  \text{max}\left[10^{-11}\left(\frac{m_X}{10^7~\TeV}\right)^{-2/3}, ~4\times 10^{-10}\left(\frac{m_X}{10^7~\TeV}\right)^{-1/2}\right] ~.
\end{align}

The first and second terms correspond to the minimal force and minimal kinetic energy criteria, respectively. Comparing Eq.~\eqref{eq:vground} (which reduces to $v_{\rm term}$ near the threshold $m_X$ for penetration) with the above, we find that a heavy particle starting from an initial height $H=10~\text{m}$ plows into the ground if 
\begin{align}
    m_X&\gtrsim 2\times 10^7\TeV &&\text{(penetrates ground from water)}~;\\
    m_X&\gtrsim 2\times 10^5\TeV &&\text{(penetrates ground from air)} ~.
\end{align}
It is easier to penetrate the ground from air because the heavy bound state has a higher terminal velocity in the air.

\subsubsection{Bouncing relaxation}
In a water reservoir with $H=10~\text{m}$, a typical CHAMP bound state with $4\times 10^{6}\TeV\lesssim m_X\lesssim 2\times 10^7\TeV$ is in the direct-fall regime, reaches the terminal velocity before reaching the ground ($t_{\rm ground}\gg t_{\rm drag}=m_X /(\rho_A a^2v_{A,\rm th})$), and bounces off the ground at the terminal speed $v_{\rm term}=v_{\rm drift}\sim 2\times 10^{-10}(m_X/10^7\TeV)$. Subsequently the motion of $X$ consists of repeated ascents and descents with energy dissipated by drag.

Equilibration requires that the drag reduce the maximum speed of $X$ from $v_{\rm term}$ to $v_{X,\rm th}=5\times 10^{-11}(m_X/10^7\TeV)^{-1/2}$, which is $\ll v_{\rm term}$.  The energy dissipated per bounce with ground-impact speed $v_{\rm max}$ is $\Delta E\sim F_{\rm drag}v_{\rm max}^2/g\sim \rho_A a^2v_{A,\rm th}v_{\rm max}^3/g$, where $v_{\rm max}^2/g$ is the maximum height for a gravity-dominated ascent (true for both $v_{\rm max}\ll v_{\rm term}$ and $v_{\rm max}=v_{\rm term}$ because $v_X=v_{\rm term}$ corresponds to $F_{\rm drag}=m_Xg$). Assuming the fractional energy dissipation per bounce is small $\Delta E/E\ll 1$ where $E\sim m_X v_{\rm max}^2$ and each bounce takes approximately $\Delta t\sim v_{\rm max}/g$, the fractional energy dissipation rate is $|\dot{E}|/E\sim \rho_A a^2v_{A,\rm th}/m_X \sim t_{\rm drag}^{-1}$. Hence, the time it takes to reach equilibration is $\sim t_{\rm drag}\ln (v_{\rm term}^2/v_{X,\rm th}^2)\sim [m_X/(\rho_A a^2v_{A,\rm th})]\ln(v_{\rm term}^2/v_{X,\rm th}^2)$. Numerically this turns out to be much shorter than the initial $t_{\rm ground}$ for $4\times 10^{6}\TeV\lesssim m_X\lesssim 2\times 10^7\TeV$. Thus, the equilibration timescale $t_{\rm eq}$ is dominated by the initial fall and is well approximated by 
\begin{align}
    t_{\rm eq}\sim t_{\rm ground}\quad (m_X\sim 4\times 10^6\text{--}2\times 10^7\TeV) ~,
\end{align}
which is given in Eq.~\eqref{eq:tground} and boils down to $\mathcal{O}(\text{minutes})$ in the aforementioned $m_X$ range.

\subsubsection{Penetration depth}
Next, we consider the case that the bound state penetrates into the ground upon impact. In solid, there is an additional energy loss channel at the level of  $dE/dx\sim \eV/\AA$ arising  from breaking chemical bonds and rupturing the lattice. The neutralized CHAMP bound state of interest is unlikely to be stripped down to a naked $X^\pm$: an $X^-$ bound to an ion $A^+$ is unlikely to lose all its electrons; and although an $X^+e^-$ bound state may lose its electron in a collision, it may rapidly recapture another electron in a conductor (e.g., tin lining), where free electrons are abundant. It is therefore reasonable to assume the effective stopping power of a slow CHAMP plowing through a solid is given by $dE/dX\sim \eV/\AA$, yielding a penetration depth,
\begin{align}
    d_{\rm plow}\sim &\frac{m_Xv_{\rm ground}^2}{\eV/\AA}\nonumber\\
    \sim& \text{min}\left[4\text{ cm}\left(\frac{m_X}{10^{10}~\TeV}\right)^3\left(\frac{\rho_A}{1~\text{g/cm}^3}\right)^{-2},\right.\nonumber\\
    &\left.0.1\text{ mm}\left(\frac{m_X}{10^{10}~\TeV}\right)\left(\frac{H}{10~\text{m}}\right)\right] ~.
\end{align}
For an impact from a water reservoir, $d_{\rm plow}$ sets the minimum thickness of the capture layer (e.g., tin lining) placed at the bottom of an enrichment tank. A 10 cm-thick tin, for example, would catch essentially all particles with $m_X\lesssim 10^{12}\TeV$.

\section{Ice cores: age versus depth}
\label{app:icecore}
\begin{figure}[t!]
    \centering
    \includegraphics[width=0.5\linewidth]{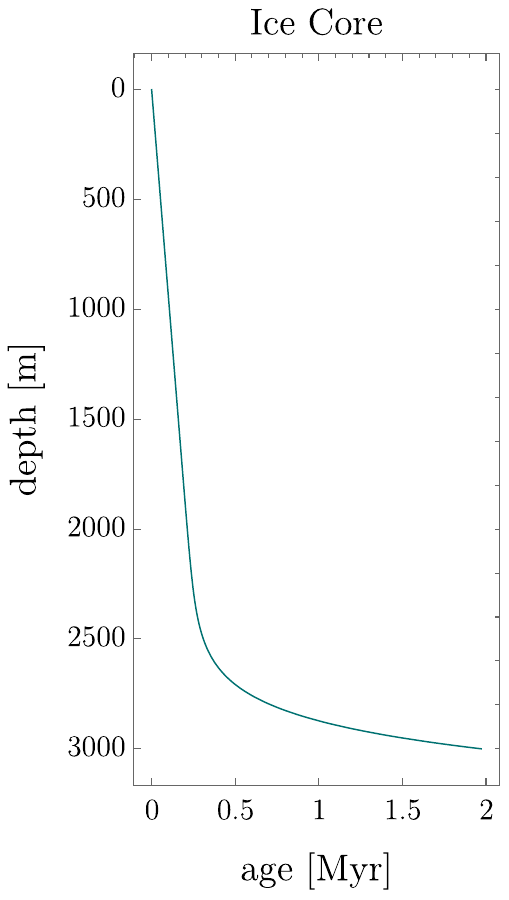}
    \caption{Benchmark ice-core depth-age relation, fitted from Ref.~\cite{IceCore2023}.}
    \label{fig:tage}
\end{figure}

Since younger ice layers are continuously deposited on top of older ones, the age $t_{\rm age}$ of an ice core increases with depth $z$. Ice-core dating typically combines multiple methods, including annual layer counting, radiometric inference using isotopes such as $^{14}\mathrm{C}$, $^{10}\mathrm{Be}$, and $^{238}\mathrm{U}$, as well as detailed glaciological modeling. We assume the following age–depth relation,
\begin{align}
    t_{\rm age}(z)&\approx t_{\rm *}\left[\frac{z}{z_*}+\frac{1}{20}\left(\frac{z}{z_*}\right)^{20}\right]~,
\end{align}
which we obtain by fitting the results of Ref.~\cite{IceCore2023} and show in Fig.~\ref{fig:tage}. Ref.~\cite{IceCore2023} speculates that the inflection at $z \sim 2600\,\mathrm{m}$ may be associated with complex flow effects related to basal melting.

\section{Enrichment strategies}
\label{app:enrichment}

{\it Liquefaction.---} Since our proposed enrichment methods require the sample to be in liquid form, liquefaction is a necessary preliminary step for solid samples. For ice-core samples, this can be achieved simply by allowing the ice to melt into water under controlled conditions. This step is straightforward, as the ice-core volume considered here is relatively modest and can be handled in a laboratory setting without significant technical challenges.

For monitored ice, a practical challenge is the efficient melting of large-volume samples. As a benchmark, consider an extraction rate of $\gtrsim{\rm (10 \, m)^3/day}$. Using the latent heat of fusion of ice, the energy required to melt this volume is approximately $\sim 3000 \, {\rm kW\cdot day}$. Assuming an industrial electricity price of $\sim \$ 0.08/{\rm kWh}$, the corresponding cost would be roughly $\$ 6000/{\rm day}$. This amounts to $\sim \$6\times 10^5$ in total for $10^{5}\text{ m}^3$ ice, suggesting that melting cost can become a practical constraint and may limit the total processable ice volume. However, this estimate assumes the most straightforward approach, direct electrical heating without optimization. The actual cost could be reduced substantially through more efficient methods, such as circulating warm or room-temperature water around the ice sample, recovering waste heat, or utilizing geothermal energy sources. A detailed optimization of the melting strategy is beyond the scope of this work and is deferred to a dedicated experimental design study.

For the geological rock samples, the strategy is to melt them and chemically convert them into a compound that is liquid at room temperature. Since most of Earth’s mantle is composed of silicon dioxide ($\mathrm{SiO}_2$)~\cite{Composition_Crust}, one possibility is to process the material into silicone oil, e.g., polydimethylsiloxane (PDMS)~\cite{PDMS} (see the discussion in Ref.~\cite{melting_LHC_2025}). The details of the required chemical processing are beyond the scope of this work. Alternatively, one could simply melt the sample by applying heat. To avoid losing extremely heavy particles, rather than fully liquefying the rock sample at once, one can gradually “broil” it so that melting proceeds from the top while a solid bottom layer is maintained to capture sinking $X$ particles. Note that these rock samples do not involve extremely large volumes (about ${\rm m^3}$) and can therefore be handled under controlled and optimized experimental conditions.

{\it Gravitational enrichment.---} In Appendix~\ref{app:eqtime}, we study the equilibration dynamics of CHAMP particles, showing that they settle to a Boltzmann height within the volume in which they are confined. Based on this geometric separation, we design a CHAMP enrichment protocol ({\it gravitational enrichment}): the liquid sample is transferred to a container and allowed to reach equilibrium; after the equilibrium timescale, only the bottom $\sim 4h_{\rm Boltz}$ layer of the container is retained, and the procedure is repeated until the sample is reduced to the desired volume for subsequent detection steps. If the container height is $H$, then at each iteration the sample volume is reduced by a factor of $\sim H/4h_{\rm Boltz}$.

For the mass range $m_X \gg 10^3\,{\rm TeV}$, the Boltzmann height is relatively small, $h_{\rm Boltz} \ll H \sim 10\,{\rm m}$, leading to efficient gravitational enrichment at each iteration. As a benchmark, consider $m_X \sim 10^4\,{\rm TeV}$, for which the Boltzmann height is $h_{\rm Boltz} \sim \mathrm{cm}\,(m_X/10^4\,{\rm TeV})^{-1}$. The detailed optimization of the enrichment procedure is mass-dependent and also depends on available resources. Here, we present a proof-of-principle example for the largest water volume considered in our proposed search. Starting from a large-volume ($10^7\,\mathrm{m^3}$) water sample, we envision the following stages: 
\begin{itemize}
    \item {Stage 1:} After the accumulation time of interest ($\sim \mathrm{yr}$), we wait for the equilibration time and then collect the bottom $4h_{\rm Boltz}$ layer of the full monitored sample of $10^7\,\mathrm{m^3}$, assuming a depth $H \sim 10\, \mathrm{m}$. For the benchmark mass $10^4\,\mathrm{TeV}$, this yields an enriched sample volume of $4\times10^4\,\mathrm{m^3}$ for the next stage. The enrichment factor at the end of this stage is ${\rm 10\,m/4\,cm}=2.5\times10^2$.
    
    \item {Stage 2:} We transfer the $4\times10^4\,{\rm m^3}$ sample into 10 containers of dimensions $10\,\mathrm{m}\times(20\,\mathrm{m})^2$ each. After waiting for the equilibration time, we collect the bottom $4h_{\rm Boltz}$ layer of each container, leaving a total of $160\,\mathrm{m^3}$ of enriched sample. The cumulative enrichment factor at the end of this stage is $({\rm 10\,m/4\,cm})^2=6.25
    \times10^4$.

    \item {Stage 3:} We transfer the enriched sample to a container of dimensions $10\,\mathrm{m} \times (4\,\mathrm{m})^2$. After equilibrium, we collect the bottom $4h_{\rm Boltz}$ layer, leaving $0.64\,\mathrm{m^3}$ of enriched sample. The cumulative enrichment factor at the end of this stage is $({\rm 10\,m/4\,cm})^3\simeq1.6\times10^7$.

    \item {Stage 4:} We transfer the enriched sample to a container of dimensions $10\,\mathrm{m} \times (25\,\mathrm{cm})^2$. After equilibrium, we collect the bottom $4h_{\rm Boltz}$ layer, leaving $2.5\times10^3\,\mathrm{cm^3}$ of enriched sample. The cumulative enrichment factor at the end of this stage is $({\rm 10\,m/4\,cm})^4\simeq4\times10^{9}$.

    \item {Stage 5:} We transfer the enriched sample to a container of dimensions $1.6\,\mathrm{m} \times (4\,\mathrm{cm})^2$. After equilibrium, we collect the bottom $4h_{\rm Boltz}$ layer, leaving $64\,\mathrm{cm^3}$ of enriched sample. The cumulative enrichment factor at the end of this stage is $({\rm 10\,m/4\,cm})^4\times(1.6\,\mathrm{m}/4\,\mathrm{cm})\simeq1.6\times10^{11}$.

    \item {Stage 6:} We transfer the enriched sample to a container of dimensions $2.56\,\mathrm{m} \times (5\,\mathrm{mm})^2$. After equilibrium, we collect the bottom $4h_{\rm Boltz}$ layer, leaving $\sim 1\,\mathrm{cm^3}$ of enriched sample. The cumulative enrichment factor at the end of this stage is $({\rm 10\,m/4\,cm})^4\times(1.6\,\mathrm{m}/4\,\mathrm{cm})\times(2.56\,\mathrm{m}/4\,\mathrm{cm})=10^{13}$, reaching the target enriched sample size.

\end{itemize}
Thus, the choice of retention length $\sim 4h_{\rm Boltz}$ leads to $(1-e^{-4})^6\sim 90\%$ of the $X$ particles remaining in the final sample.

As shown in Fig.~\ref{fig:timescale}, the equilibration timescale for $m_X \gtrsim 10^3\,{\rm TeV}$ is $t_{\rm eq}\sim 20\,{\rm days}~(m_X/10^3\,{\rm TeV})^{-1}$, assuming convection is negligible. Therefore, with the handful of enrichment stages outlined above, the total enrichment time remains below the year-scale duration that we consider feasible. This required time rapidly decreases for larger masses ($\sim$ month for $m_X \sim 10^4\,{\rm TeV}$), since both the Boltzmann height becomes smaller and the equilibration time shortens.

For smaller masses, however, the total enrichment time increases approximately as $\propto m_X^{-(n+1)}$, where $n$ is the number of stages that require waiting to reach equilibrium. As a result, relying solely on gravitational enrichment quickly becomes impractical below $m_X \sim 10^3\,{\rm TeV}$ while maintaining an initial sample volume of $10^7\,{\rm m^3}$. In this lower-mass regime, we instead limit the initial sample size to the volume that can be processed via centrifugal enrichment (see discussion below), namely $\sim 10^5\,{\rm m^3}$. This transition is reflected in Fig.~\ref{fig:reach} as a sharp change in sensitivity around $m_X \sim 10^3\,{\rm TeV}$.

In the higher-mass regime considered in this work, CHAMPs can have microscopic Boltzmann heights. A potential concern is that such particles may become trapped or adhere to container walls, whose surface roughness can be comparable to the Boltzmann height. To mitigate this loss mechanism, we propose lining the container walls (especially the bottom) with a low-melting-point metal (e.g., tin). During subsequent enrichment stages, this surface-layer lining can be melted and enriched itself.

{\it Centrifugal enrichment.---}
Centrifugal enrichment is an established method for separating high-mass components within a sample. Some of us analyzed centrifuge-based enrichment for CHAMP particles at the lower end of the mass range considered in this work in Ref.~\cite{melting_LHC_2025}, where we argued that volumes as large as $10^5\,{\rm m^3}$ can be processed efficiently. We refer the reader to Ref.~\cite{melting_LHC_2025} for further details. Based on that analysis, we set the projected reach in the lower-mass regime ($m_X \lesssim 10^3\,{\rm TeV}$) by requiring $N_X \gtrsim 1$ within a $10^5\,{\rm m^3}$ sample.

More generally, centrifugal enrichment serves as a complementary tool within our overall enrichment strategy and can be implemented at various stages, or applied to different target samples, for logistical optimization. For example, gravitational and centrifugal enrichment can be combined to reduce the total processing time. This is particularly relevant around $m_X \sim 10^3\,{\rm TeV}$, where we assumed a sharp transition from gravitational to centrifugal enrichment, effectively limiting the initial sample volume by a factor of 100. A hybrid strategy could smooth this transition and potentially extend the reach in this mass range.

\bibliography{references}

\end{document}